\def\arxivversion{1}
\documentclass[letterpaper,twoside]{article}

\usepackage[top=2.5cm, bottom=2.5cm, left=2.0cm, right=2.0cm]{geometry}
\usepackage{graphicx}
\DeclareGraphicsExtensions{.pdf,.png,.jpg,.mps,.eps,.ps}
\graphicspath{{figs/}{.}}
\usepackage[numbers,sort&compress,super,comma]{natbib}
\usepackage[table]{xcolor}
\usepackage{booktabs} %
\usepackage{multirow}
\usepackage{enumitem}
\definecolor{darkblue}{rgb}{0.05,0.05,0.4}
\usepackage{hyperref} %
\hypersetup{
    colorlinks = true,
    citecolor = darkblue,
    linkcolor = darkblue,
    }
\usepackage{orcidlink} %

\newif\ifsecnumbers
\secnumberstrue
\newcommand{\secref}[1]{%
  \ifsecnumbers Sec.~\ref{#1}\else Sec.~\nameref{#1}\fi}
\newcommand{\secrefpair}[2]{%
  \ifsecnumbers Sec.~\ref{#1} and~\ref{#2}\else Sec.~\nameref{#1} and~\nameref{#2}\fi}
\newcommand{\twosectionsref}[2]{%
  \ifsecnumbers Sections~\ref{#1} and~\ref{#2}\else Sections~\nameref{#1} and~\nameref{#2}\fi}

\usepackage{url}
\usepackage{tikz}
\usetikzlibrary{arrows.meta, positioning, fit, backgrounds, shapes.geometric, calc}
\usepackage[most]{tcolorbox}

\newtcolorbox[auto counter,number format=\Alph]{explainbox}[2][]{%
    colframe=gray!60!black,
    colback=gray!5,
    coltitle=white,
    fonttitle=\bfseries,
    sharp corners=south,         %
    rounded corners=southeast,   %
    boxrule=0.8pt,               %
    drop shadow=black!20,        %
    title=Box~\thetcbcounter: #2, #1,
    before upper={\setlength{\parindent}{1.5em}}, %
}

\usepackage{microtype} %
\usepackage[T1]{fontenc} %
\usepackage[default]{lato} %
\renewcommand{\normalsize}{\fontsize{8.2pt}{10pt}\selectfont}
\normalsize
\usepackage{fancyhdr}
\usepackage{titling}
\usepackage{stfloats} %
\usepackage{caption} %
\DeclareCaptionFont{sevenpt}{\fontsize{7pt}{8.5pt}\selectfont}
\pretitle{\begin{center}\LARGE}
\posttitle{\end{center}}
\preauthor{\begin{center}
\large \lineskip 0.5em}
\postauthor{\par\end{center}
\begin{center}
\textit{
$^1$Champalimaud Centre for the Unknown, Champalimaud Foundation, Lisbon, Portugal\\
$^2$Janelia Research Campus, Howard Hughes Medical Institute, Ashburn, VA, USA\\
$^3$Harvard University, Cambridge, MA, USA\\
$^4$Yale University, New Haven, CT, USA\\
$^5$University of Cambridge, Cambridge, United Kingdom\\
$^6$\'Ecole Polytechnique F\'ed\'erale de Lausanne (EPFL), Lausanne, Switzerland\\
$^7$RyvivyR Inc, NY, USA\\
$^\dagger$Co-first authors\qquad $^\ddagger$Co-senior authors\\
$^\ast$Corresponding authors: \textnormal{\texttt{\{memming.park, gonzalo.polavieja, alfonso.renart, joe.paton\}@research.fchampalimaud.org}}
}
\end{center}}

\title{
Integrative neurocybernetic modeling
\\
in the era of large-scale neuroscience
}
\author{
I.~Memming~Park$^{1,7,\dagger,\ast}$\,\orcidlink{0000-0002-4255-7750},
Gonzalo~G.~de~Polavieja$^{1,\dagger,\ast}$\,\orcidlink{0000-0001-5359-3426}, %
Ayesha~Vermani$^{1}$, %
Juan~\'Alvaro~Gallego$^{1}$, %
Kathleen~N.~Esfahany$^{3}$, %
Shreya~Saxena$^{4}$, %
Guillaume~Hennequin$^{5}$, %
Michael~Orger$^{1}$, %
Auke~Ijspeert$^{6}$, %
Matthew~Dowling$^{1,7}$, %
Daniel~McNamee$^{1}$, %
Srinivas~C.~Turaga$^{2}$, %
Zachary~Mainen$^{1}$, %
Alfonso~Renart$^{1,\ddagger,\ast}$\,\orcidlink{0000-0001-7916-9930}, %
Joseph~J.~Paton$^{1,\ddagger,\ast}$\,\orcidlink{0000-0002-1663-6455} %
}
\date{}

\usepackage{amssymb}
\usepackage{mathtools} %
\usepackage{bm} %
\usepackage[mathlines,switch]{lineno}

\usepackage{letltxmacro}
\LetLtxMacro{\originaleqref}{\eqref}
\renewcommand{\eqref}{Eq.~\originaleqref}
\definecolor{c:adjointidx}{rgb}{0.15,0.55,0.7}
\definecolor{c:time}{rgb}{0.50,0.12,0.7}

\usepackage{forloop}
\newcommand{\defvec}[1]{\expandafter\newcommand\csname v#1\endcsname{{\mathbf{#1}}}}
\newcounter{ct}
\forLoop{1}{26}{ct}{
    \edef\letter{\alph{ct}}
    \expandafter\defvec\letter
}

\forLoop{1}{26}{ct}{
    \edef\letter{\Alph{ct}}
    \expandafter\defvec\letter
}

\definecolor{mpcolor}{rgb}{1, 0.1, 0.59}

\ifsecnumbers
\else
\fi

\usepackage[compact]{titlesec}
\titlespacing*{\section}{0pt}{4pt}{0pt}
\titlespacing*{\subsection}{0pt}{4pt}{0pt}

\begin{document}
\ifdefined\arxivversion\else\linenumbers\fi
\maketitle
\thispagestyle{fancy}

\begin{abstract}
Large-scale neuroscience is generating rich datasets across animals,
brain areas, and behavioral contexts, yet our modeling efforts remain fragmented across isolated experiments.
We argue that understanding behavior requires integrative neurocybernetic models:
understandable dynamical models that capture the closed-loop coupling of brain,
body, and environment, treat the brain as a controller pursuing latent objectives,
represent structured variation across scales, and scale to heterogeneous datasets.
Such models shift the goal from predicting neural recordings in isolation
to inferring the organizing principles that govern neural and behavioral dynamics.
We outline a practical route toward this goal by combining nonlinear state-space models
and meta-dynamical extensions with scalable inference, knowledge distillation, and
mixed open- and closed-loop training.
By pooling complementary constraints from recordings, behavior, perturbations, and anatomy,
integrative neurocybernetic models can provide statistical amplification,
few-shot generalization, and mechanistic insight into shared dynamical structure,
individual variation, and the control objectives that govern behavior.
This agenda offers a model-centric path from fragmented data to a mechanistic science of how brains produce behavior.
\end{abstract}

\twocolumn

\section{Introduction}
Massive neural recordings across animals, behaviors, and environments
hold promise for data-driven discovery of the organizing principles of the brain,
yet the field lacks modeling tools that can integrate across fragmented experiments and extrapolate to a broader context.
Recent efforts to scale neural data analysis have borrowed the engineering playbook of artificial intelligence (AI),
but scaling alone does not produce scientific understanding.

Behavior emerges from intricate feedback loops linking neural circuits, the body, environmental dynamics, and interactions with other agents (\secref{sec:neurocybernetic}).
How, then, can experiments and modeling reveal the neural basis of this cybernetic, or control-systems, view of behavior?
In an ideal, though unrealistic, experimental paradigm, we could
imagine recording large-scale, high-quality measurements of all relevant variables
from a well-controlled subject across a broad range of environmental contexts.
Although recent progress has been made in this direction for simple
model organisms such as \emph{C.~elegans} and larval zebrafish,
whose neural circuits are well characterized
and whose responses are relatively
stereotyped~\cite{Haspel2023-if,simeon2024homogenized,Brennan2019-rc,Lueckmann2024-db},
realizing this ideal is still challenging and understanding behavior from it is fundamentally unattainable: %
the space of behaviors and environments that a brain
has evolved to handle---shaped by the cumulative experience of countless ancestors---cannot be adequately
sampled from a single animal in one experimental lifetime.

In reality, neural and behavioral data are collected from various
noisy, heterogeneous, non-simultaneous, and independent experiments with many hidden and uncontrolled factors.
Putting together each piece of noisy evidence in a way that would facilitate the construction of a coherent theory of behavior
remains a daunting task for a single research group and, more practically, a major statistical challenge.
The consequences are both technical and scientific:
statistical barriers to integrating heterogeneous evidence
effectively constrain which questions neuroscientists can practically ask,
biasing research away from questions about organizing principles
toward narrower characterizations of individual recordings from particular brain regions in limited contexts.

While this divide-and-conquer paradigm has yielded many valuable insights into subsystems of the brain and body, it cannot capture the behavior of an organism. The continuous interaction between neural subsystems, body, and environment gives rise to behaviors that no isolated subsystem predicts: central pattern generators are modulated by descending commands and peripheral sensory feedback; motor plans are reshaped in real time by proprioceptive signals~\cite{Fujiwara2022-oh,Cruz2023-bd,Ijspeert2023-fg,Dickinson2000-cu}; arousal states emerge from brainstem--cortex loops that have no meaningful description outside their coupled operating regime~\cite{Brookings2012-ia}. Data collected from an isolated subsystem are still useful---they provide statistical constraints on the full model---but we currently lack an efficient means of integrating data from individual tasks, regions, and recordings into a coherent whole.

Recent research efforts have drawn inspiration from AI and machine learning (ML), where scaling in data, computation, and collective innovations have unlocked unprecedented performance.
These have primarily pointed neuroscience toward establishing benchmarks\cite{Pei2021-kf,Sun2023-sa,Karpowicz2024-qe} and datasets\cite{Teeters2008-mv}, and toward developing foundation model architectures (Box~\ref{box:foundation}) adapted for neural and behavioral datasets.
While these developments have been promising, and there is growing interest in large-scale models in the field, one critical question emerges:
how do we direct our research efforts on large-scale modeling toward the primary scientific objective of understanding biological neural systems and their behavior?

In this Perspective, we outline a path forward for such an integrative framework for neuroscience
that leverages the benefits of scaling while remaining grounded in
scientific discovery and understanding (\secref{sec:understandability}).
Specifically, we argue for a model class that can embody a holistic, cybernetic theory,
in addition to capturing the statistical regularities across diverse neural and behavioral datasets.
Such integrative models can
(1) provide statistical amplification by combining constraints from
multiple sources (\secref{sec:statamp}), and
(2) systematize shared structures across individuals (and species), forming a foundation for neuroscientific discovery.

We argue that integration should take place at three levels.
First, models should integrate across {\em function}, being able to describe a multiplicity of tasks and, in sum, general-purpose behavior in real time.
Second, integration should take place across {\em brain systems}, as behavior is the result of their interaction.
Third, progress will only be made if we are able to integrate across {\em datasets}, with the model functioning as a nexus that brings together knowledge and constraints derived from a wide range of partial records of brain-behavior interactions.
We believe that a community-driven effort to build a shared integrative model of neural and embodied behavior,
\textit{an integrative neurocybernetic model}, would transform systems neuroscience,
turning isolated empirical findings into a coherent understanding of
the principles governing brain function.

\section{Desiderata for the integrative model}

The types of models we seek
must do more than fit data (Fig.~\ref{fig:integrative-objectives}).
They must be \emph{understandable}, supporting explanation rather than prediction alone;
and \emph{neurocybernetic}, capturing the closed-loop interplay among brain, body, and environment.
They must be capable of representing \emph{structured variation across scales}, so that shared principles and meaningful differences across contexts, individuals, and species can be studied within a common framework\cite{Grillner2020-zl}.
They must also be \emph{scalable}, so that they can integrate heterogeneous datasets at the scope demanded by modern neuroscience. This set of properties defines the goals of the scientific program we advocate for in this Perspective.

\begin{explainbox}[label={box:foundation},float=t]{{Foundation model for neuroscience}}
    Foundation models (FMs)\cite{Bommasani2021} have two defining characteristics:
    (1) an FM learns the joint statistical structure of the data through large-scale pretraining, and
    (2) the learned representations allow FMs to solve a variety of \textbf{downstream tasks}.
    Therefore, the downstream tasks determine the usefulness of foundation models.

    In systems neuroscience, FMs must learn a common representation of neural dynamics across labs, species, experimental setups, and tasks,
    so that a new dataset can be rapidly adapted \emph{few-shot} instead of training a bespoke model from scratch.
    Notably, recent efforts\cite{Azabou2023-on,Azabou2024-bs,Ryoo2025-hl,Ye2023-za,Vermani2024b,Zhang2025-im}
    report evidence for \textbf{cross-session, -animal, and -species generalization},
    rapidly adapting to unseen subjects with minimal labels,
    indicating that broad pretraining can transfer across primates, rodents, artificial recurrent neural networks (RNNs), and humans.
    To date, attempts at building FMs for neuroscience still have
    large gaps due to heterogeneities in and limited access to data 
    (cf.~\cite{Orhan2025-sg}), alignment (\secref{sec:alignment}), training objectives, and lack of scientifically meaningful downstream benchmarks.
    Moreover, most of them have been developed primarily for decoding
    (predicting behavior or task variables from neural activity) or
    diagnostic classification, rather than building a model
    that can dynamically generate ethologically relevant behavior (Box~\ref{box:neurocybernetic}).
    The performance of current FMs in neuroscience is still far from that observed in FMs for natural language or image processing.

    FMs for neuroscience should consider downstream tasks that align with
    the scientific end goals of systems neuroscience to understand
    how the brain and body work in diverse contexts and give rise to flexible behavior (Fig.~\ref{fig:integrative-objectives}).
    If advances in AI are to meaningfully inform neuroscience,
    they must be designed to include the inherently dynamical nature of biological neural systems\cite{Rust2025-et,Vermani2024b,Vermani2024a}.
\end{explainbox}

\subsection{Understandability}\label{sec:understandability}
For scientific purposes, models must ultimately support understanding rather than mere prediction,
favoring forms of modeling that expose assumptions, mechanisms, and causal structure
in ways that human researchers can actively reason with~\cite{Woodward2004-gb,Kaplan2011-uw}.

Foundation models for neuroscience (Box~\ref{box:foundation}) are %
powerful pattern learners
which hold the promise to 
become valuable
as instruments for large-scale data analysis, simulation, and hypothesis generation\cite{Dyer2025-tt} (\secref{sec:alignment}).
However, their internal representations are, at present,
only weakly aligned with the conceptual frameworks
in which scientific explanation and theory building actually occur\cite{Momennejad2023-du,Mathis2025-tn}.

In contrast, more interpretable models---%
whether mechanistic, causal, or deliberately simplified statistical constructions---%
can expose structure in forms that scientists can reason about.
Their understandability may arise from the possibility of mapping model components onto biological parts
or information-processing computations, or from cognitively friendly abstractions.
Interpretable models enable articulation of {\em why} a result holds in terms of 
how interventions would alter it, and where the model fails when confronted with new evidence~\cite{Woodward2004-gb}.
For example, a recurrent circuit model of decision-making with distinct excitatory and inhibitory populations
can be informative not only because it reproduces behavioral and neural data,
but because one can ask how changing recurrent gain, noise, or inhibition alters evidence accumulation,
and use failures under novel perturbations to diagnose limits of the proposed mechanism~\cite{Wong2006-ey,Jha2024-wn}.
\textbf{A large black-box model might predict the same outcomes more accurately,
but in a mature scientific workflow such models should inform interpretable ones,
not replace them as the primary locus of explanation, critique, and
theory formation.}
This tension between prediction performance and scientific understanding is
central to current debates over foundation models for neuroscience\cite{Dyer2025-tt}.

\subsection{Neurocybernetics}\label{sec:neurocybernetic}

An agent maintains an internal state distinct from the external world,
acts on the environment through a body, receives sensory feedback,
and generates actions. While the ultimate role of normativity in biology is contested~\cite{lynch2007frailty, silver2021reward, marken2021study, moreno2026intrinsic}, we find it useful to make the modeling assumption that actions are guided by implicit control objectives with the ultimate goal of survival, and we refer to the agent's actions viewed in this context as ``meaningful behavior''.

In this sense, the integrative model must be able to \textbf{behave}, above and beyond its ability to encode or decode neural recordings well. That is, when coupled with a body and an ethologically relevant dynamical environment,
it should operate as a causal dynamical system that generates meaningful behavior in closed loop (Box~\ref{box:cybernetic})\cite{Monaco2024-wn,Vaxenburg2025-fr,Haimerl2025-uj}.
We call this the \textbf{neurocybernetic} desideratum.
Recent work shows that both system dynamics and control objectives can be inferred from partial observations of closed-loop behavior~\cite{Kwon2020-uh,Geadah2024-hv},
underscoring the feasibility of identifying neurocybernetic models from data (\secref{sec:ssm}). We call a model of the nervous system that structurally resembles and operates as such a closed-loop embodied agent a \textbf{neurocybernetic} model (Box~\ref{box:neurocybernetic}).

Autoregressive Transformer-style models in robotics offer an instructive contrast:
they generate behavior moment-to-moment from past observations\cite{Reed2022-tm,Sun2023-sa,Brohan2022-kb,Zitkovich2023-zh} and
have shown promising behavior-cloning results; however, their internal mechanics neither resemble biological processes nor map naturally onto the dynamical models used to describe those processes in neuroscience.
For example, the large context window and the global injection of
temporal order and distance information used in Transformer architectures are biologically implausible as a dynamical mechanism.
More broadly, any black-box approach that reproduces behavioral sequences without causal dynamical structure
necessarily falls short of the neurocybernetic desideratum.
We require that the model's internal dynamics correspond to
biologically identifiable components---brain regions, cell types, neuromodulatory systems---and
that behavior arises from their causal interaction in closed loop to
serve (implicit) control objectives.
Notably, despite being non-neurocybernetic,
black-box architectures such as Transformers are effective at scaling (\secref{sec:scalability}) and can still guide less trainable neurocybernetic models (\secref{sec:kd}).

\begin{explainbox}[label={box:cybernetic}]{Cybernetic model}
    \begin{center}
        \includegraphics[width=3in]{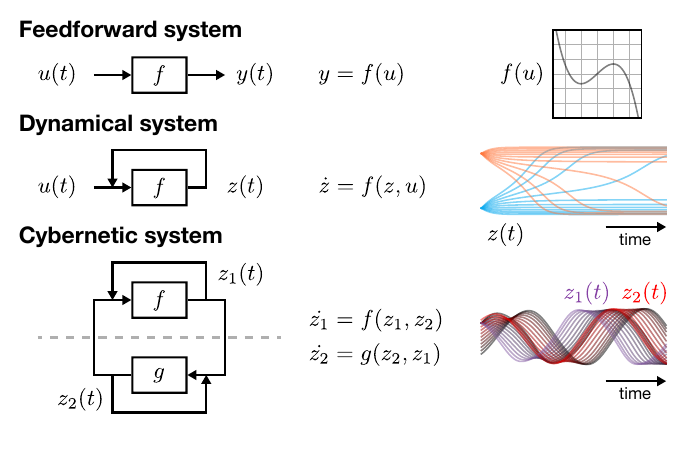}
    \end{center}
    A simple feedforward system maps inputs to outputs while lacking an internal state.
    Instead of studying a system only as an input-output mapping, a dynamical approach introduces a feedback loop and defines an internal state $z(t)$.
    A cybernetic system further separates the agent and environment, each as a stateful dynamical system interacting in a feedback loop.
    Mathematically, coupling one dynamical system to another defines a new, larger dynamical system.
    Only in the context of the coupled system can we observe their often new and emergent dynamical behavior.
    What distinguishes a cybernetic system from a generic coupled dynamical system is that the agent is a \emph{controller}:
    it maintains a latent control objective and acts to regulate its sensory
    input\cite{Powers1973-zu} and internal state accordingly (\secref{sec:neurocybernetic}).
\end{explainbox}

\subsection{Structured variation across scales}\label{sec:structVar}
Not all variation in neural data is created equal. Some reflects universal principles shared across contexts, individuals, and species; some reflects scientifically meaningful differences, such as individual strategies, developmental trajectories, learning histories, and neurological conditions; and some is transient noise or recording artifact. These sources of variation occur at different scales. Within an individual, neural dynamics change with context, attention, arousal, neuromodulatory state, and satiation\cite{IBL2025-il,IBL2021-ii,Gutierrez2013-hr}.
Across individuals, they vary with development, learning, genetic background, body morphology, and pathology.
Variation across species introduces another scale. Morphology and behavioral repertoires can diverge substantially while aspects of neural circuit organization and control remain conserved over large evolutionary distances~\cite{Grillner2018-limb,Katz2013-circuits,Grillner2020-zl}.

Because we lack comprehensive tools for integrating data across different sources of variation, variability across individuals, for instance, is generally averaged out~\cite{pagan2025individual}, precluding both a systematic investigation of differences in individual behavior and a principled way of learning what is common across them.
Even across species there might be shared organizing principles that we would like to understand and harness for efficient learning.
The integrative model should therefore allow common circuit motifs, dynamical computations, or control principles to be shared when supported by the data, while retaining (i.e., explicitly modeling) differences in behavioral strategy, neural implementation, and even morphology. By pooling empirical constraints across individuals, sessions, and experimental conditions, an integrative model gains the statistical power (\secref{sec:statamp}) to resolve structure that would otherwise be inaccessible in any single dataset~\cite{friston2001dynamic,Kiebel2008-db,sussillo2014neural}.

\begin{explainbox}[label={box:neurocybernetic}]{Neurocybernetic models}
    \begin{center}
        \includegraphics[width=2.5in]{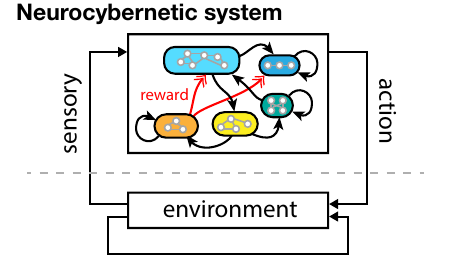}
    \end{center}
    Neurocybernetics studies the nervous system as an embodied agent
    with control objectives (Box~\ref{box:cybernetic}).
    A neurocybernetic model generates behavior both in response to changes in the environment
    and autonomously due to internal state changes.
    The nested feedback loops at multiple anatomical and temporal
    scales can be studied neurocybernetically.
    The pioneers of cybernetics, Norbert Wiener\cite{Wiener1965-xv} and Ross Ashby\cite{Ashby1952-lt},
    envisioned such an explanation of brain function in the 1940--1960s,
    and William Powers made the control of perception, rather than of action,
    the organizing principle of behavior\cite{Powers1973-zu};
    but the meaning of `cybernetics' has since broadened to many other disciplines;
    we re-introduce the original concept with a more specific
    terminology, \emph{neurocybernetics}.
\end{explainbox}

In our view, a hierarchical generative architecture provides a natural way to model structured variation, with higher levels corresponding to gradual changes that parameterize families of lower-level processes that modulate faster changes (\secref{sec:meta-ssm}). Under this modeling approach, what is learned is not a single model, %
but a \textbf{structured family of models}, organized so that meaningful variation at each scale is {\em explicitly represented}, captured by interpretable factors and separated from noise.
As we have seen, there are naturally separating timescales of changes with corresponding biophysical mechanisms of change.
Transient changes---context shifts, cognitive state fluctuations---correspond
to rapid excursions that wash out within the behavioral timescale,
while slow, persistent changes---learning, development, disease---trace
extended trajectories at higher levels.
Evolution and species-level differences define the broadest boundaries of the family.

Of course, in addition to providing a natural solution to the representation of structured variation, hierarchical architectures are a prominent feature of nervous systems.
Motor control provides a particularly clear illustration: lower-level circuits (spinal reflexes, central pattern generators, brainstem motor programs) supply
relatively automatized movement primitives\cite{Michaels2020-xs}.
These modules may be able to operate largely independently, yet together form a compositional repertoire of movement building blocks that can be flexibly recruited and coordinated under higher-level control\cite{Ijspeert2013-jq,Cruz2023-bd,Kashtan2005-me,Schimel2024-mm}.
This hierarchical decomposition dramatically simplifies the high-level control problem, a principle independently rediscovered in modern hierarchical reinforcement learning~\cite{Merel2020-ly}.

Indeed, modularity and compositionality are also generally observed in systems organized hierarchically, including brains. Neural systems such as the basal ganglia, cortex, and cerebellum exhibit distinct circuit architectures and functional specializations, and are themselves composed of repeated subnetworks performing related
cybernetic computations on different inputs\cite{Yang2019-ja,Haimerl2025-uj}.
Recent studies have begun to investigate how recurrent neural networks learn compositional strategies when asked to perform groups of tasks that have a hierarchical structure~\cite{driscoll2024flexible,riveland2024natural,Lazzari2025-jo}.
Other works have focused on the conditions under which modular architectures emerge~\cite{Dobs2022-ff,Schug2024-sc,Dorrell2025-pv,Lange2022-wc}, including cases where modularity appears to be driven by the need to solve multiple hierarchically organized tasks~\cite{wu2026task}.
Discovering such hierarchical and parallel modular organizations directly from data, and relating them to known anatomy while modeling the full system jointly, would constitute a major advance.
We note that hierarchical, modular systems solving problems compositionally are expected to produce robust structured variation, driven by alternative ways of combining motifs and modules in pursuit of functional goals~\cite{kirschner1998evolvability,csete2002reverse}.

In \twosectionsref{sec:ssm}{sec:meta-ssm},
we propose state-space models and their meta-dynamical extensions as a concrete realization of this hierarchical generative framework.

\subsection{Scalability}\label{sec:scalability}

Successful models that capture the behavior of organisms holistically are necessarily going to be large and complex, and constraining those models is inevitably going to require very large amounts of data. Progress in FMs (especially large language models) has undoubtedly been driven by the scaling up of both model parameters and training data\cite{henighan2020scaling}.
Similarly, progress in neuroscience now depends on the ability to learn from large-scale, high-quality datasets that span multiple animals, brain areas, and behavioral contexts.
We believe that scalability will be essential for the success of such an integrative modeling approach\cite{Sutton2019-bl}. Recent studies show that the performance of neuroscience Transformers tends to saturate with model size but not with data size\cite{gokce2025scaling,willeke2026omnimouse,gokce2026multimodal}, highlighting current limitations in the quantity and diversity of datasets in neuroscience. Architectures incorporating better inductive biases suffer less in the data-limited regime\cite{gokce2026multimodal}, supporting the idea that model architectures capturing essential design elements of the brain are not only desirable in terms of interpretability and understanding, but can also support more sample-efficient training\cite{bordelon2022population}.

A crucial ingredient necessary for scaling is 
\emph{computational efficiency}: models must be trained and deployed within realistic resource budgets. The Transformer architecture dominates modern AI not for biological plausibility, but because it scales gracefully with existing hardware, enabling unprecedented performance at scale\cite{Sun2025-lz}.
Several foundation models for neuroscience follow this direction, hoping that emergent capabilities useful for advancing neuroscientific understanding will arise at scale. By contrast, current dynamical models for neural data either prioritize expressivity at the expense of efficiency or rely on inference algorithms that are too slow or fragile to handle the scale and heterogeneity of contemporary neuroscience experiments.

What is missing is a class of models and methods that are both dynamically structured and computationally scalable.
State-space models provide an interpretable framework for neural dynamics,
but current training strategies require further innovations to reach the scale required for modern data, and to exploit hardware acceleration as effectively as Transformer-based methods. Conversely, Transformer and diffusion-style models scale well but lack the inductive biases and expressive power necessary for mechanistic interpretation of neural computation. Bridging this gap is an open problem: we must seek high-performance inference algorithms that can ingest and learn from large-scale, heterogeneous datasets while retaining interpretability and theoretical grounding. Knowledge distillation (\secref{sec:kd}) is a promising approach to achieve this goal.

\subsubsection{Statistical amplification}\label{sec:statamp}

Statistical amplification is the power of an integrative model to leverage shared structure across diverse datasets,
enabling the extraction of reliable patterns that might be too subtle or rare to detect in any single experiment.
By lifting the statistical barriers inherent to isolated experiments,
integrative models expand the space of scientific questions the field can meaningfully ask.
Integrative modeling should not only enhance statistical sensitivity,
but also help generalize findings to new subjects and contexts, facilitate few-shot learning,
aid in online experimental designs\cite{Jeon2026a},
and allow for principled imputation or prediction in the face of missing data.
Even in the absence of interpretability, statistical amplification provides a pragmatic advantage,
serving as a statistical prior/posterior that improves inference and discovery throughout the neuroscience modeling pipeline.
An integrative model should function as such a prior, fulfilling a primary promise of neural foundation models\cite{Zhang2025-im}.
In addition to scaling, statistical amplification is also essential for representing structured variation (\secref{sec:structVar}) and as such can shed new light on difficult problems such as the mechanistic consequences of learning: asking whether we can track changes in effective neural dynamics from \textit{only the recorded activity} reframes the problem as a tractable statistical question that benefits from pooling across individuals and sessions.

\begin{figure*}[b!]
\centering
\includegraphics[width=0.9\textwidth]{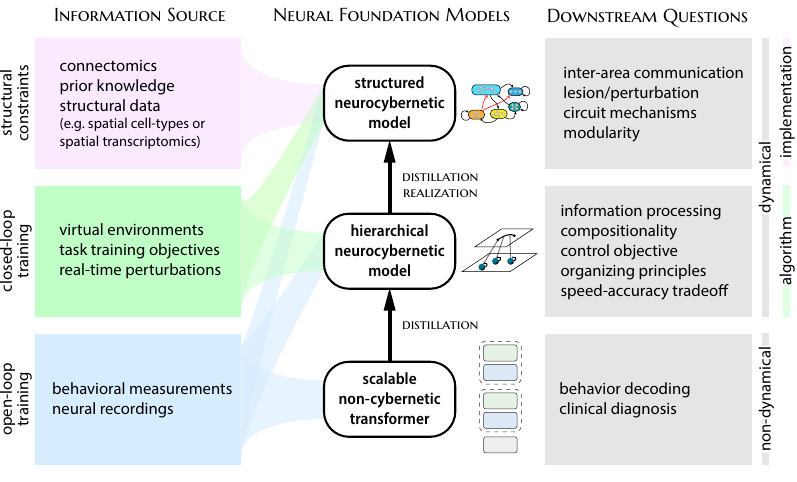}
\caption{
\textbf{Extending integrative neural foundation models from prediction to neurocybernetic mechanism.}
Scalable, non-cybernetic Transformer models extract some statistical structures
from open-loop neural and behavioral recordings (Box~\ref{box:foundation}),
but prediction is only a beginning.
To move from prediction toward mechanism,
we propose training scalable models in closed loop (\secref{sec:mixed-training})
and constraining them with biological structure (\secref{sec:connectomics})
to construct hierarchical and structured neurocybernetic models.
Knowledge distillation (\secref{sec:kd})
and biological realization connect these model classes.
The downstream questions shown are illustrative rather than exhaustive,
ranging from behavioral decoding and diagnosis
to cross-species organizing principles and circuit mechanisms.
}
\label{fig:integrative-objectives}
\end{figure*}

\subsubsection{Challenges: Alignment %
}\label{sec:alignment}
A significant challenge to the scaling up needed in neural FMs is the problem of data alignment.
To train an FM, we need to align data from diverse sources into a common representational space to extract the underlying statistical structure.
In natural language processing and computer vision, this alignment is straightforward: text in a given language is composed of a discrete set of characters, and images share consistent pixel grids, which can be used to construct standardized features.
In contrast, experimental data in systems neuroscience present a unique set of challenges that prevent straightforward alignment and unified training.
Even when the latent dynamics that one would like to capture are conserved\cite{Gallego2020-lj,Safaie2023-pq,Pandarinath2018,Kaifosh2025-rn}, neural responses are highly variable, and demonstrate drift in tuning properties over time.
Moreover, each experiment gives us a partial, non-overlapping subsample of neural population activity, often during entirely different behavioral contexts.
In practice, the problem is that an additional layer is needed in each experiment to learn the mapping from the idiosyncratic group of neurons in each recording to the latent factors performing putatively conserved computations (and similarly for different output effectors).
The question is then whether, as the amount of data grows, sufficient learning of the shared computational backbone will occur or whether most of the parameters and learning budget will go into learning these per-session, per-subject alignment layers.
The scaling behavior across sessions of predictions of mouse neural activity and even short-term forecasting was recently evaluated and consistent increases were observed up to a few hundred sessions\cite{willeke2026omnimouse}, suggesting that shared learning is still taking place at these scales.
How these trends will look when many thousands of sessions from more heterogeneous experiments are used is still an open question.

In recent years, two primary approaches have been used to address this alignment challenge:
(1) learning embeddings for individual units while leveraging cross-attention~\cite{Azabou2023-on}, and
(2) learning separate alignment functions for each recording session~\cite{Pandarinath2018}.
An important next step is to compare how these alignment strategies scale with recording size and dataset heterogeneity and whether functional cell types or overcompleteness in the recordings would provide benefits in data efficiency.

\section{Promising approaches}

Based on the desiderata, we outline a possible roadmap toward a framework for integrative modeling (Fig.~\ref{fig:integrative-objectives}).
In the first stage, a scalable, non-cybernetic predictive model (Box~\ref{box:foundation})\cite{Azabou2024-bs,Zhang2025-im,willeke2026omnimouse,duan2025poco} would be trained on recordings of sensory input, behavior, and neural activity to predict masked neural and behavioral data.
In the second stage, this predictive model would serve as the teacher (\secref{sec:kd}) for distilling a neurocybernetic hierarchical state-space model that represents shared dynamics and structured variation (\secrefpair{sec:ssm}{sec:meta-ssm}).
The resulting model would interact with the environment in real time and learn to meet control objectives while remaining constrained by recorded data (\secref{sec:mixed-training}).
It could be used to identify dynamical principles and algorithms underlying behavior.
In the third stage, a second distillation step would train a neurocybernetic model incorporating structural constraints such as modules resembling brain systems, individual brain areas, or cell types %
(\secref{sec:connectomics}).
Such a model would let us ask questions about neural implementation.
The building blocks we discuss next each address a distinct facet of integrative modeling---from representational structure to scalability to closed-loop behavior.
Together, they provide \textbf{a path toward progressively more informative models}, rather than toward a single ideal model that recovers the underlying biological system completely.
Neural and behavioral data provide only partial and heterogeneous constraints on that system, and multiple dynamical explanations may remain consistent with the available evidence.
Method development should therefore focus on extracting the scientifically meaningful structure these constraints can support, while making uncertainty and remaining ambiguity explicit.
Existing approaches address different parts of this roadmap, but none yet provides such an integrative framework on its own.

\subsection{Nonlinear state-space models}\label{sec:ssm}
To achieve large-scale neural dataset integration, we must choose the
right level of abstraction---one that preserves biological fidelity
while providing understandability and generalization.
A latent neural dynamical systems view provides such a level\cite{van-Gelder1998,Rust2025-et,Driscoll2018-lw}.
Rather than relying on raw electrophysiological or optophysiological recordings or their summary statistics, we build a structured representation, i.e., state-space models (SSMs) that capture how the \emph{latent neural state evolution over time} explains the observed time series.

\begin{figure*}[t!b!]
\centering
\vspace{3mm}
\includegraphics[width=0.7\textwidth]{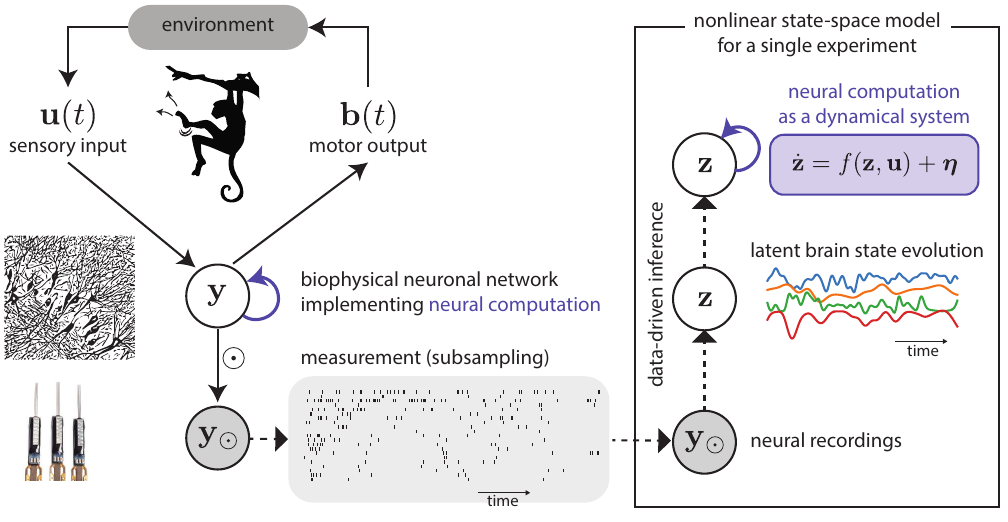}
\caption{\textbf{State-space modeling as a framework for neural
dynamics.} State-space models explain observed neural and
behavioral time series through latent dynamical states and an
observation model. The central inferential task is to recover the
underlying nonlinear dynamics from partial, noisy measurements
(\secref{sec:ssm}).}
\label{fig:SSM}
\end{figure*}

In an SSM, the high-dimensional neural recordings are explained by the evolution of neural states that capture the coordinated activity patterns,
and the dynamical systems that govern how these trajectories evolve.
The statistical challenge is to learn the unknown neural states and dynamical systems from only the observations (Fig.~\ref{fig:SSM}).

We can formalize the inference problem as follows:
Given partial neural/behavioral $\{\vy_t\}$ and partial input $\{\vu_t\}$ time series, can we build a concise state-space model that captures the underlying nonlinear dynamics responsible for their generation~\cite{Haykin1998}?
More specifically, we would like to identify a continuous nonlinear process that captures the temporal structure, and an instantaneous noisy observation process:
\begin{align}
    \mathrm{d}\vz &= f(\vz(t), \vu(t))\,\mathrm{d}t + \mathrm{d}\bm{\eta}_t \qquad &\text{(state dynamics)}
    \label{eq:dynamics}
    \\
    \vy(t) \mid \vz(t) &\sim P(\vy \mid g(\vz(t), \vu(t))) \qquad &\text{(observation model)}
    \label{eq:obs}
\end{align}
where $\mathrm{d}\bm{\eta}_t$ is the process/state noise that captures unobserved perturbations of the state $\vz$, $g$ and $f$ are continuous functions, and $P$ denotes a probability distribution for the noisy observation~\cite{Sarkka2013,Roweis2001}.
SSMs provide an interpretable framework for complex time series analysis
by combining an intuitive dynamical system model with a probabilistic observation model.
Critically, the function $f(\cdot)$ is particularly important because it encodes the latent dynamics through which the model represents neural computation.
In neural data analysis, state-space models played a key role in providing insights into neural dynamics~\cite{Breakspear2017,Kao2015b,Paninski2009,Yu2009,Ecker2014b}, neural computation~\cite{Mante2013,Zhao2016d,Nair2025-rj}, development of neural prosthetics and treatment through feedback control~\cite{ODoherty2011,Gilja2012-so,Little2012,Willett2021-dp,Hocker2019a},
and control objectives during closed-loop behavior~\cite{Geadah2024-hv,Kao2021-xg}.

However, as mentioned in \secref{sec:scalability}, flexible nonlinear SSMs often present challenges in terms of trainability and scaling.
Recent advances in scalable variational inference for exponential-family observations have made nonlinear SSM fitting tractable on large neural datasets~\cite{Dowling2024b,Pandarinath2018}.
Moreover, nonlinearly stacked linear SSMs~\cite{Gu2024-jj,Dowling2026a,Yang2024-vi,Smith2022-dy}
have gained popularity as an innovative alternative to the Transformer architectures to deal with the poor scaling with context length
(i.e., the quadratic bottleneck) of traditional attention mechanisms,
offering a more scalable foundation for modeling long-range dependencies.
They approximate nonlinear dynamics by composing fast linear blocks, taking advantage of modern massively parallel hardware through the associative scan algorithm~\cite{sarkka2020temporal}.
Beyond stacked linear approximations, recent work has shown that even genuinely nonlinear state-space models can be evaluated in parallel by recasting the state sequence as a fixed-point problem solved via Newton's method~\cite{Gonzalez2025-nm,Gonzalez2025-yw,Danieli2025-jw}.
We believe we should harness these advances, along with numerical acceleration strategies~\cite{Gander2008-uu},
to bridge nonlinear SSMs with fast, stacked-linear generative and inference models.

Even if these methods permit more efficient model fitting, optimizing an independent SSM for each dataset does not by itself explain how the resulting dynamical systems relate across animals, tasks, or contexts, nor does it allow us to aggregate evidence to constrain the common dynamical motifs across datasets (\secref{sec:structVar}).
Doing so requires a representation not only of neural state, but of variation across dynamical systems themselves.

\subsection{Meta-dynamical state-space models}\label{sec:meta-ssm}
\begin{figure*}[t!b!]
    \centering
    \includegraphics[width=0.8\textwidth]{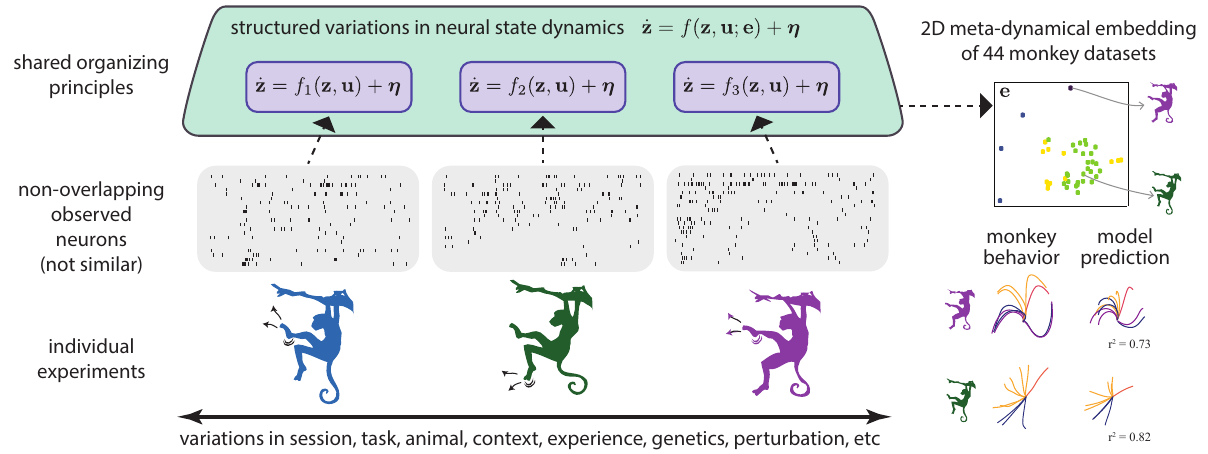}
\caption{
Integrative model at the level of meta-dynamical system (modified from \citet{Vermani2024b}).
Note that the neural recordings themselves cannot be aligned or compared in general.
The complex yet shared dynamics along with the structured variations capture the organizing principle of neural computation.
(\secref{sec:meta-ssm})
}
\label{fig:metaDS}
\end{figure*}

To capture both the shared structure and the systematic variations in neural computation across contexts, \citet{Vermani2024b} introduced a new concept: \textbf{the meta-dynamical space} (Fig.~\ref{fig:metaDS}).
This space defines a family of neural dynamical systems, where each point corresponds to a specific instance of dynamics\cite{Cotler2023-kr,Linderman2019-ng} capable of explaining a particular neural recording and associated behavior (Fig.~\ref{fig:metaDS}).

We can think of it as adding another layer of hierarchy to the SSM defined as \eqref{eq:dynamics} and \eqref{eq:obs}.
\begin{align}
f(\vz(t), \vu(t)) &\sim P(f \mid \ve)
\end{align}
where $\ve$ denotes the meta-dynamical space embedding that captures the particular ``context'' for the current dataset (\secref{sec:structVar}).

They showed that a pretrained meta-dynamical state-space model\cite{Vermani2024b} learns a map from each dataset to a 2-dimensional representation
capable of instantiating animal-, task-, and condition-specific dynamical SSMs while exposing the shared structure for reaching-type behavior in monkeys (Fig.~\ref{fig:metaDS} right).
Notably, the model enabled few-shot generalization to new animals: orders of magnitude fewer trials than the typical thousands sufficed to instantiate a reasonable nonlinear dynamical model\cite{Vermani2024b}.

At least two open challenges remain.
First, the meta-dynamical embedding is currently learned in a purely data-driven manner, so the axes of variation it discovers need not correspond to scientifically meaningful factors;
encouraging interpretable structure in the embedding---for instance through disentanglement objectives or supervision from known experimental covariates---is an important next step.
Second, the current parameterization treats the dynamical family as a monolithic mapping from embedding to dynamics;
ideally, it should encourage anatomically modular or compositional structure, so
that changes along one axis of the meta-dynamical space correspond to
changes in a specific subsystem (e.g., a motor module or a sensory
pathway) rather than a global reconfiguration---without restricting the mapping to circuit-specific correspondences.
How to impose such compositionality without sacrificing flexibility is an open problem on its own~\cite{Schug2024-sc,Riveland2026-cosyne}.
A promising route comes from decomposed linear dynamical systems
(dLDS)\cite{Yezerets2025-by} that express nonstationary neural
dynamics as time-varying combinations of a shared dictionary of linear
operators, recovering context-dependent ``dynamic connectivity''
within an individual \emph{C.~elegans} and a latent space aligned across
individuals. %

\subsection{Knowledge distillation}\label{sec:kd}
The understandability desideratum (\secref{sec:understandability}) posits that the modeling pipeline must ultimately produce representations with genuine systematic understandability.
Knowledge distillation (KD) is a mechanism that can realize this goal through a multi-stage pipeline:
it transfers what a scalable but opaque model has learned into a structured model that scientists can reason with,
interrogate through interventions, and incrementally refine.

There is a gap between what a model can theoretically express (expressivity) and what it can learn to do in practice (trainability).
Models that are flexible enough to express the desired function often find only inferior solutions through training;
this gap is larger for the smaller and more structured models and is known as the optimization error~\cite{Bottou2007-ge}.
KD bridges this gap by using a more trainable teacher model to indirectly obtain a structured student model~\cite{Ba2014-fl,Hinton2015-qe,Fournier2023-ed}.
For example, the teacher's knowledge is transferred by encouraging the student's internal representations and outputs to match the teacher through a differentiable objective.

The choice of student architecture determines the form and nature of the resulting scientific knowledge (Fig.~\ref{fig:integrative-objectives}).
The spectrum of possible students ranges from highly symbolic to highly flexible:
at one end, sparse symbolic regression methods such as SINDy~\cite{Brunton2016-oa} distill dynamics into closed-form equations
that resemble physics-like descriptions, maximizing interpretability at the cost of representational flexibility;
in the middle, flexible nonlinear ODEs and RNNs retain expressive power while providing a dynamical systems object amenable to analysis;
at the other end, restricted structures such as switching linear dynamical systems~\cite{Smith2021} or low-rank RNNs~\cite{Valente2022-hr}
expose the geometry and stability of neural computations through analytically tractable forms.
In neuroscience, such model simplification has a rich history.
Fitting an expressive RNN or SSM and identifying dynamical features such as
fixed points, slow manifolds, and bifurcation structure~\cite{Smith2021,Mante2013,Genkin2020-ps,Schaeffer2020-ib,sussillo2013opening,Sagodi2025b}
is a type of knowledge distillation for scientific interpretability.
What distinguishes formal KD from these earlier approaches is that it provides a differentiable and automatable objective,
applicable beyond the case-by-case theoretical analyses that have driven the field so far.
We emphasize that, unlike KD in ML where the goal is computational efficiency, our goal is to increase understandability:
the student model should expose assumptions, mechanisms, and causal structure that support scientific reasoning.

\begin{explainbox}[label={box:mixed-training}]{Mixed open/closed-loop training}
    \begin{center}
        \includegraphics[width=2.5in]{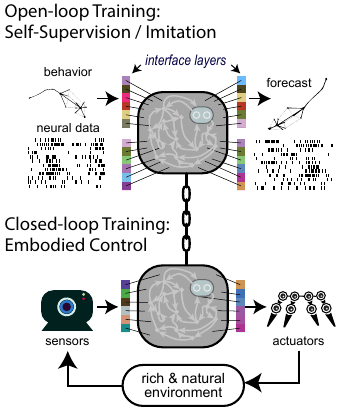}
    \end{center}
    \textbf{Open-loop training} fits the model to pre-recorded trajectories; the model does not influence the environment.
    \textbf{Closed-loop training} deploys the model as an embodied agent whose actions influence future inputs through environmental feedback.
    A mixed training protocol links the constraints from both training methods to combine data efficiency with feedback-dependent adaptation (\secref{sec:mixed-training}).
\end{explainbox}

A concrete and promising instantiation is to use autoregressive Transformer models as teachers and structured RNNs as students.
Transformers can be efficiently trained using teacher forcing on a self-supervised objective, since they do not maintain hidden states.
Biologically structured RNNs, in contrast, suffer from trainability issues including vanishing and exploding gradients and high computational cost for backpropagation through time.
In \citet{Xia2025-am}, a Transformer-based model was used to interpolate missing joint information in neural and behavioral recordings, paving the way for training an RNN via KD.
More recently, \citet{Kumar2026-kr} proposed a novel approach to
training RNNs using a Transformer encoder on a predictive-state
objective that provides the desired RNN hidden states without backpropagation through time.

An interesting question is whether distillation could be geared explicitly toward
scientific interpretability---and not simply for higher benchmark values. An open challenge is defining fidelity criteria that go beyond output accuracy to extract dynamical structure, causal relationships, and modular organization that may be only implicit and incomplete in the teacher.
Once distilled, the resulting structured model can be further refined through closed-loop interaction with the environment (\secref{sec:mixed-training}) and through data-driven fine-tuning.

\subsection{Mixed open/closed-loop training}\label{sec:mixed-training}
Training integrative neurocybernetic models requires learning from
both passive observational data and active interaction with
ethologically relevant environments.
Neither purely off-policy (open-loop) nor purely on-policy (closed-loop) training is sufficient for robust, generalizable behavior implemented by the brain (Box~\ref{box:mixed-training}).

Off-policy methods, such as behavior cloning and system identification from passive recordings, enable efficient learning from demonstrations and existing datasets but suffer from covariate shift: models trained on fixed datasets fail when deployed because the agent's own actions lead to states not well represented in the training distribution~\cite{Ross2011-aa}. Conversely, purely on-policy methods are sample-inefficient and potentially unstable.
More fundamentally, even with accurate body emulation, purely on-policy learning usually does not lead to natural animal-like behavior because we lack access to the internal objectives, reward functions, and normative principles that guide animal behavior---complex internal drives including homeostatic needs, social motivations, and evolved behavioral strategies that are difficult to specify a priori~\cite{Vaxenburg2025-fr,Levine2020-aa}.
In addition, the simulated body and environment must approximate the organism's morphology and ecological niche,
so that the feedback dynamics shaping learned behavior match those that shaped the real nervous system~\cite{Brunton2026-sphinx}.

Integrative behavioral models must combine both:
respecting structure captured by passive observational data while being shaped by closed-loop interaction to capture feedback-dependent dynamics\cite{Aldarondo2024-zb}.
A practical training protocol may proceed in phases:
(1) pretrain model components in open loop on diverse recorded data to learn shared representations,
(2) gradually introduce closed-loop embodied behavior episodes with safety constraints, and
(3) jointly optimize for both open-loop and closed-loop objectives, while optionally constraining closed-loop behavior toward the offline distribution for stability.
Joint optimization ties the model parameters learned in one regime
with the behavior of the model in the other regime (Box~\ref{box:mixed-training}).
This strategy incentivizes the alignment of the dynamic core across both open- and closed-loop settings. 

Mixed open/closed-loop training connects to offline reinforcement learning with online fine-tuning~\cite{Levine2020-aa}, imitation learning with on-policy refinement (e.g., DAgger~\cite{Ross2011-aa}), system identification followed by adaptive control,
and joint inference of system dynamics and control objectives from closed-loop observations,
where non-identifiability from passive data alone makes active exploration fundamentally necessary~\cite{Geadah2024-hv,Vermani2024a}.
The resulting neurocybernetic model is constrained to both explain observed data and generate meaningful behavior as an embodied agent.

\subsection{Structurally constrained %
models}\label{sec:connectomics}
Brains are characterized by strong structure-function relationships.  Peripheral sensory systems, the cerebral cortex, cerebellum, hippocampus, basal ganglia, and the superior colliculus each have highly structured architectures both locally and in terms of the way they are connected to each other. A tradition in neuroscience has tried to explore how the function of these systems is specifically related to their architecture \cite{marr1969theory,treves1994computational,sparks1988neural,rao2024sensory}. In addition to the activity- and behavior-related constraints that we have emphasized so far, the integrative modeling framework should also be capable of accommodating structural constraints. 

Existing approaches include linking latent states to the activity of specific neurons and imposing connectivity and sign constraints (to model excitatory and inhibitory interactions) in the matrix describing the interaction between the latents \cite{Jha2024-wn,karniol2024modeling,liu2025accurate}, treating neural activity from individual neurons within the SSM as causal elements in the dynamics, rather than just observations \cite{linderman2016bayesian,Valente2022-hr,pals2024inferring}, or modifying learning algorithms so that they automatically give rise to structured networks \cite{balwani2025constructing}.

These, or newly developed approaches, could be incorporated into the integrative framework by combining structural constraints as another level of refinement to SSMs distilled from a large-scale black-box model (Fig.~\ref{fig:integrative-objectives}). Through a second distillation step, one can progressively include anatomical or cell-type information while simultaneously satisfying behavioral and neural activity records learned as described above.
Closer correspondence between the causal structure of an integrative FM and anatomical and biophysical details can help identify how components of the brain and body work together.
It can link population-level function to circuit mechanisms and the dynamical roles of cell types and connectivity~\cite{Lappalainen2024-np}, predict responses to perturbations and guide experiments~\cite{pospisil2024fly}, and describe structured variation at the level of cells and circuits~\cite{Schlegel2024-multiconnectome}.

The rapid progress in connectomics offers a powerful source of structural constraints which are being harnessed for the development of models of brain function\cite{Lappalainen2024-np,Zhao2024-baaiworm,Pugliese2025-va,beiran2025prediction}.
In some sense, a connectome---with its implied full use of all neural elements as causal determinants of the dynamics---lies at the opposite end of the modeling spectrum from the usual dimensionality reducing SSMs, in which the role of the state is to compactly summarize unmeasured parts of the full dynamical system.
However, connectome-constrained SSMs have been used to generate biophysical models of C-elegans\cite{Mi2022-iclr}, treating the activity of each neuron as a latent variable.
Alternatively \citet{pospisil2024fly} used the drosophila connectome as a prior to infer the causal effect of single neurons (the `effectome') in the fly using optogenetic perturbations. In a similar spirit, connectomes within the integrative framework constrain the space of plausible dynamical models rather than uniquely specifying one.
Demonstrating that connectomes can be directly used as a structural basis of interpretable student models in knowledge distillation (\secref{sec:kd})---offering a practical route to training biologically structured RNNs that may otherwise be difficult to fit directly---would be an important advance in this direction.

\section{Discussion}\label{sec:discussion}

Systems neuroscience needs integrative neurocybernetic models: closed-loop dynamical systems coupling brain, body, and environment,
designed to be understandable rather than seeking post-hoc explanations from a black-box model.
The primary objective is understanding---how neural dynamics give rise to behavior---and
the goal is not merely to predict neural activity but to build behaving agents whose internal structure reflects the biology that produced them.
Such models can serve as a shared substrate where heterogeneous experiments accumulate constraints and statistical power amplifies across the community.
Our proposal is to utilize the strength of several approaches to build progressively more useful integrative FMs
(Fig.~\ref{fig:integrative-objectives}).

\subsection{From aspiration to program}
The cybernetic vision of brain function is not new---Wiener~\cite{Wiener1965-xv} and Ashby~\cite{Ashby1952-lt} articulated it decades ago.
What has changed is the deep learning revolution and the broader scaling program in machine learning.
GPU-accelerated differentiable programming, self-supervised pretraining at scale, foundation models trained on heterogeneous data~\cite{Bommasani2021}, and the empirical demonstration that capability gains are predictable from compute and data~\cite{Sutton2019-bl} have together produced an ecosystem of methods that did not exist when cybernetics fell out of fashion.
The specific building blocks we have emphasized---scalable state-space architectures~\cite{Gu2024-jj,Dowling2026a}, knowledge distillation~\cite{Hinton2015-qe}, and validated mixed offline/online training protocols~\cite{Levine2020-aa,Ross2011-aa}---are only a small slice of what can now be brought to bear on neuroscience problems.
Together, these advances turn the cybernetic program from a philosophical aspiration into a concrete engineering and scientific enterprise---much as general AI escaped its repeated winters once sufficient compute, data, and algorithms converged.

\subsection{From large-scale efforts to integration}
Large-scale neuroscience is no longer a dream.
The International Brain Laboratory has demonstrated multi-lab standardized recordings at unprecedented scale~\cite{IBL2021-ii,IBL2025-il}.
The Allen Institute provides whole-brain atlases and systematic surveys of cell types and connectivity.
MICrONS has produced dense connectomic reconstructions of mammalian cortex.
Several neural foundation models---Transformer-based architectures, inpainting models, and self-supervised representation learners~\cite{Azabou2023-on,Ye2023-za,Ryoo2025-hl,Zhang2025-im}---have shown that cross-session and cross-animal generalization is feasible.

What is missing is the collective push toward integration.
Each of these efforts focuses on a particular slice of the problem---standardized behavior, anatomical completeness, or representational transfer---without synthesizing them into a unified generative model of behaving organisms.
Repetitive data of the same type provide diminishing returns; what amplifies statistical power is complementary information that jointly constrains the model from multiple angles.

\subsection{Data demands and limits}
Although in principle an integrative model should be able to absorb any data, the datasets that will be more useful will be those that span across multiple constraining dimensions simultaneously. This can take the form of brain and behavior, multiple brain areas simultaneously, the same brain area across multiple tasks, etc. Similarly, electrophysiology, imaging, behavior, connectomics, and perturbations can each reduce ambiguities left by the others, provided they sufficiently overlap in the biological processes they constrain. More data of the same type eventually provide diminishing returns; what matters is adding independent and informative constraints. Since we view behavior as deriving from the satisfaction of hidden control objectives, neural activity from %
brain areas believed to encode such goals would be particularly useful \cite{cisek2019resynthesizing,theriault2025s,Nair2025-rj}. In terms of behavior, more complex tasks away from the overtrained regime are associated with less redundant neural activity across the brain and thus will be more useful to identify the unique role of different regions\cite{gao2015simplicity,khilkevich2024brain}.
More quantitatively, scaling-law analyses could help quantify how much different forms of data contribute by measuring how model performance and generalization change not only with dataset and model size~\cite{Kaplan2020-scaling,Antonello2023-scaling,gokce2025scaling,willeke2026omnimouse,gokce2026multimodal}, but also as training expands across animals, conditions, modalities, species, and perturbations.

That said, instead of a dedicated platform for experimental collection under prespecified conditions, an essential goal of the integrative program is to be flexible enough to leverage the many existing, weakly coordinated data collection efforts from the whole neuroscientific community. Even then, integration does not guarantee a uniquely identifiable model. Multiple dynamical or mechanistic explanations may remain consistent with the available evidence. The goal should therefore be to progressively constrain the space of plausible models, identify what remains uncertain, and use those uncertainties to guide the next experiments. In this way, integrative modeling can make data collection increasingly informative.

\subsection{Toward a community-scale program}
Building an integrative neurocybernetic model will require sustained collaboration among experimentalists, theorists, and computational neuroscientists,
each contributing expertise the others cannot substitute.
Progress must be anchored in continuous validation.
We envision a benchmarking ecosystem---analogous to what has driven ML progress---%
but with evaluation criteria aligned to neurocybernetic modeling:
not just prediction accuracy on held-out neural data, but closed-loop behavioral fidelity,
few-shot generalization to new animals and conditions, recovery of known dynamical structure in ground-truth settings,
and the interpretability of discovered organizing principles.

\section*{Acknowledgements}
We would like to thank everyone who participated in the discussions during the Neuro-cybernetics at Scale symposium in October 2025 at the Champalimaud Foundation, Lisbon, Portugal.
The authors used Anthropic Claude for editorial assistance; all scientific content, arguments, and final wording are the authors' responsibility.

\bibliographystyle{plainnat_memming_v1}
\bibliography{refs,catniplab}

@Unpublished{Jeon2026a,
    author = {Hyungju Jeon and Il Memming Park},
    title = {Beyond the attractor: closed-loop stimulation design for learning multistable neural dynamics},
    journal = {bioRxiv},
    year = {2026},
    month = oct,
    doi = {10.64898/2026.10.05.756678},
    url = {https://www.biorxiv.org/content/10.64898/2026.10.05.756678v1},
  themes = {neurotech},
  tags = {neural-control;real-time;active-learning;state-space-models},
  note = {bioRxiv preprint, doi: 10.64898/2026.10.05.756678},
}

@Unpublished{Dowling2026a,
    author = {Matthew Dowling and Hyungju Jeon and Cristina Savin and Il Memming Park},
    title = {Memory by Design: Probabilistic Sequence Layers},
    year = {2026},
    archiveprefix = {arXiv},
    eprint = {2605.31163},
    url = {http://arxiv.org/abs/2605.31163},
    primaryclass = {stat.ML},
  tags = {state-space-models;machine-learning},
  note = {arXiv preprint arXiv:2605.31163},
}

@InProceedings{Vermani2024b,
  title         = "Meta-dynamical state space models for integrative neural data analysis",
  author        = "Vermani, Ayesha and Nassar, Josue and Jeon, Hyungju and
                   Dowling, Matthew and Park, Il Memming",
  booktitle = {International Conference on Learning Representations (ICLR)},
  month     = apr,
  year          =  2025,
  url		= "https://openreview.net/forum?id=SRpq5OBpED",
  archivePrefix = "arXiv",
  primaryClass  = "stat.ML",
  eprint        = "2410.05454",
  themes = {inference;integrative},
  tags = {state-space-models;motor-control;transfer-learning},
  selected = {true},
}

@Unpublished{Sagodi2025b,
  title         = "Dynamical archetype analysis: Autonomous computation",
  author        = "S\'agodi, \'Abel and Park, Il Memming",
  month         =  jul,
  year          =  2025,
  url           = "http://arxiv.org/abs/2507.05505",
  archivePrefix = "arXiv",
  primaryClass  = "math.DS",
  eprint        = "2507.05505",
  themes = {language},
  tags = {continuous-attractors;dynamical-archetypes;working-memory;recurrent-neural-networks},
  selected = {true},
  note = {arXiv preprint arXiv:2507.05505},
}

@InProceedings{Vermani2024a,
  author    = {Ayesha Vermani and Matthew Dowling and Hyungju Jeon and Ian Jordan and Josue Nassar and Yves Bernaerts and Yuan Zhao and Van Vaerenbergh, Steven and Il Memming Park},
  booktitle = {European Signal Processing Conference},
  title     = {Real-time machine learning strategies for a new kind of neuroscience experiments},
  year      = {2024},
  archivePrefix = "arXiv",
  primaryClass  = "q-bio.NC",
  eprint        = "2409.01280",
  themes = {neurotech},
  tags = {real-time},
}

@InProceedings{Dowling2024b,
  title         = "{eXponential} {FAmily} Dynamical Systems ({XFADS}): Large-scale nonlinear Gaussian state-space modeling",
  author        = "Dowling, Matthew and Zhao, Yuan and Park, Il Memming",
  booktitle     = {Advances in Neural Information Processing Systems (NeurIPS)},
  month         =  dec,
  year          =  2024,
  archivePrefix = "arXiv",
  primaryClass  = "stat.ML",
  eprint        = "2403.01371",
  arxivid       = "2403.01371",
  doi		= "10.52202/079017-0430",
  url           = "https://openreview.net/forum?id=Ln8ogihZ2S",
  code		= "https://github.com/catniplab/xfads",
  themes = {inference},
  tags = {state-space-models;variational-inference},
  selected = {true},
}

@Article{Hocker2019a,
  author    = {David Hocker and Il Memming Park},
  journal   = {{PLOS} Computational Biology},
  title     = {Myopic control of neural dynamics},
  year      = {2019},
  month     = mar,
  doi       = {10.1371/journal.pcbi.1006854},
  PMID = {30856171},
  PMCID = {6428347},
  url       = {https://www.biorxiv.org/content/early/2017/12/30/241299},
  themes = {neurotech},
  tags = {neural-control;brain-disorders;real-time},
}

@InProceedings{Zhao2016d,
  author        = {Zhao, Yuan and Park, Il Memming},
  title         = {Interpretable Nonlinear Dynamic Modeling of Neural Trajectories},
  booktitle     = {Advances in Neural Information Processing Systems (NIPS)},
  year          = {2016},
  archiveprefix = {arXiv},
  eprint        = {1608.06546},
  primaryclass  = {q-bio.QM},
  youtube	= {https://www.youtube.com/watch?v=7oWRZRpaq_I},
  url = {https://papers.nips.cc/paper/6543-interpretable-nonlinear-dynamic-modeling-of-neural-trajectories},
  themes = {inference},
  tags = {state-space-models},
}

@inproceedings{karniol2024modeling,
  title={Modeling state-dependent communication between brain regions with switching nonlinear dynamical systems},
  author={Karniol-Tambour, Orren and Zoltowski, David and Diamanti, E Mika and Pinto, Lucas and Brody, Carlos and Tank, David and Pillow, Jonathan},
  booktitle={International Conference on Learning Representations},
  volume={2024},
  pages={27010--27030},
  year={2024}
}

@article{liu2025accurate,
  title={Accurate identification of communication between multiple interacting neural populations},
  author={Liu, Belle and Sacks, Jacob and Golub, Matthew D},
  journal={Proceedings of machine learning research},
  volume={267},
  pages={39381},
  year={2025}
}

@article{pals2024inferring,
  title={Inferring stochastic low-rank recurrent neural networks from neural data},
  author={Pals, Matthijs and Sa{\u{g}}tekin, A Erdem and Pei, Felix and Gloeckler, Manuel and Macke, Jakob H},
  journal={Advances in Neural Information Processing Systems},
  volume={37},
  pages={18225--18264},
  year={2024}
}

@article{linderman2016bayesian,
  title={Bayesian latent structure discovery from multi-neuron recordings},
  author={Linderman, Scott and Adams, Ryan P and Pillow, Jonathan W},
  journal={Advances in neural information processing systems},
  volume={29},
  year={2016}
}

@article{balwani2025constructing,
  title={Constructing biologically constrained RNNs via Dale’s backpropagation and topologically informed pruning},
  author={Balwani, Aishwarya and Wang, Alex Q and Najafi, Farzaneh and Choi, Hannah},
  journal={Science Advances},
  volume={11},
  number={50},
  pages={eadw4970},
  year={2025},
  publisher={American Association for the Advancement of Science}
}

@article{pospisil2024fly,
  title     = "The fly connectome reveals a path to the effectome",
  author    = "Pospisil, Dean A and Aragon, Max J and Dorkenwald, Sven and
               Matsliah, Arie and Sterling, Amy R and Schlegel, Philipp and Yu,
               Szi-Chieh and McKellar, Claire E and Costa, Marta and Eichler,
               Katharina and Jefferis, Gregory S X E and Murthy, Mala and
               Pillow, Jonathan W",
  journal   = "Nature",
  publisher = "Springer Science and Business Media LLC",
  volume    =  634,
  number    =  8032,
  pages     = "201--209",
  month     =  oct,
  year      =  2024,
  url       = "http://dx.doi.org/10.1038/s41586-024-07982-0",
  doi       = "10.1038/s41586-024-07982-0",
  pmc       = "PMC11446844",
  pmid      =  39358526,
  issn      = "0028-0836,1476-4687",
  language  = "en"
}

@article{beiran2025prediction,
  title={Prediction of neural activity in connectome-constrained recurrent networks},
  author={Beiran, Manuel and Litwin-Kumar, Ashok},
  journal={Nature neuroscience},
  volume={28},
  number={12},
  pages={2561--2574},
  year={2025},
  publisher={Nature Publishing Group US New York}
}

@article{marr1969theory,
  title={A theory of cerebellar cortex},
  author={Marr, David},
  journal={The Journal of physiology},
  volume={202},
  number={2},
  pages={437--470},
  year={1969},
  publisher={Wiley Online Library}
}

@article{sparks1988neural,
  title={Neural cartography: sensory and motor maps in the superior colliculus},
  author={Sparks, David L},
  journal={Brain, Behavior and Evolution},
  volume={31},
  number={1},
  pages={49--56},
  year={1988},
  publisher={S. Karger AG Basel, Switzerland}
}

@article{rao2024sensory,
  title={A sensory--motor theory of the neocortex},
  author={Rao, Rajesh PN},
  journal={Nature neuroscience},
  volume={27},
  number={7},
  pages={1221--1235},
  year={2024},
  publisher={Nature Publishing Group US New York}
}

@article{treves1994computational,
  title={Computational analysis of the role of the hippocampus in memory},
  author={Treves, Alessandro and Rolls, Edmund T},
  journal={Hippocampus},
  volume={4},
  number={3},
  pages={374--391},
  year={1994},
  publisher={Wiley Online Library}
}

@article{khilkevich2024brain,
  title={Brain-wide dynamics linking sensation to action during decision-making},
  author={Khilkevich, Andrei and Lohse, Michael and Low, Ryan and Orsolic, Ivana and Bozic, Tadej and Windmill, Paige and Mrsic-Flogel, Thomas D},
  journal={Nature},
  volume={634},
  number={8035},
  pages={890--900},
  year={2024},
  publisher={Nature Publishing Group UK London}
}

@article{gao2015simplicity,
  title={On simplicity and complexity in the brave new world of large-scale neuroscience},
  author={Gao, Peiran and Ganguli, Surya},
  journal={Current opinion in neurobiology},
  volume={32},
  pages={148--155},
  year={2015},
  publisher={Elsevier}
}

@article{theriault2025s,
  title={It’s not the thought that counts: Allostasis at the core of brain function},
  author={Theriault, Jordan E and Katsumi, Yuta and Reimann, Henning M and Zhang, Jiahe and Deming, Philip and Dickerson, Bradford C and Quigley, Karen S and Barrett, Lisa Feldman},
  journal={Neuron},
  volume={113},
  number={24},
  pages={4107--4133},
  year={2025},
  publisher={Elsevier}
}

@article{cisek2019resynthesizing,
  title={Resynthesizing behavior through phylogenetic refinement},
  author={Cisek, Paul},
  journal={Attention, perception, \& psychophysics},
  volume={81},
  number={7},
  pages={2265--2287},
  year={2019},
  publisher={Springer}
}

@inproceedings{duan2025poco,
  title={{POCO}: Scalable neural forecasting through population conditioning},
  author={Duan, Yu and Chaudhry, Hamza and Ahrens, Misha B and Harvey, Christopher and Perich, Matthew G and Deisseroth, Karl and Rajan, Kanaka},
  booktitle={Advances in Neural Information Processing Systems},
  volume={38},
  pages={23631--23669},
  year={2025},
  doi={10.52202/085713-0798}
}

@article{bordelon2022population,
  title={Population codes enable learning from few examples by shaping inductive bias},
  author={Bordelon, Blake and Pehlevan, Cengiz},
  journal={Elife},
  volume={11},
  pages={e78606},
  year={2022},
  publisher={eLife Sciences Publications, Ltd}
}

@inproceedings{gokce2026multimodal,
  title={Multimodal Scaling Laws for Task \& Data-Optimized Models of Visual Cortex},
  author={Gokce, Abdulkadir and Tang, Yingtian and Schrimpf, Martin},
  booktitle={International Conference on Machine Learning},
  volume={2026},
  pages={},
  year={2026}
}

@inproceedings{willeke2026omnimouse,
  title={OmniMouse: Scaling properties of multi-modal, multi-task Brain Models on 150B Neural Tokens},
  author={Willeke, Konstantin F and Turishcheva, Polina and Gilbert, Alex and Chakrabarty, Goirik and Bedel, Hasan and Fahey, Paul and Qiu, Yongrong and Weis, Marissa and Vystr{\v{c}}ilov{\'a}, Michaela and Muhammad, Taliah and others},
  booktitle={International Conference on Learning Representations},
  volume={2026},
  pages={134382--134411},
  year={2026}
}

@inproceedings{gokce2025scaling,
  title={Scaling laws for task-optimized models of the primate visual ventral stream},
  author={Gokce, Abdulkadir and Schrimpf, Martin},
  booktitle={Proceedings of the 42nd International Conference on Machine Learning (ICML)},
  series={PMLR},
  volume={267},
  pages={19712--19733},
  year={2025},
  url={https://proceedings.mlr.press/v267/gokce25a.html}
}

@article{kirschner1998evolvability,
  title={Evolvability},
  author={Kirschner, Marc and Gerhart, John},
  journal={Proceedings of the National Academy of Sciences},
  volume={95},
  number={15},
  pages={8420--8427},
  year={1998},
  publisher={The National Academy of Sciences}
}

@article{csete2002reverse,
  title={Reverse engineering of biological complexity},
  author={Csete, Marie E and Doyle, John C},
  journal={science},
  volume={295},
  number={5560},
  pages={1664--1669},
  year={2002},
  publisher={American Association for the Advancement of Science}
}

@article{wu2026task,
  author    = {Wu, Yuhang and Deng, Shikuang and Du, Kangrui and Mattar, Marcelo G. and Wu, Yifan and Bassett, Dani S. and Tang, Huajin and Pan, Gang and Gu, Shi},
  title     = {Task-structured modularity emerges in artificial networks and aligns with brain architecture},
  journal   = {Nature Machine Intelligence},
  year      = {2026},
  month     = {sep},
  doi       = {10.1038/s42256-026-01306-9},
  url       = {https://www.nature.com/articles/s42256-026-01306-9}
}

@article{riveland2024natural,
  title={Natural language instructions induce compositional generalization in networks of neurons},
  author={Riveland, Reidar and Pouget, Alexandre},
  journal={Nature Neuroscience},
  volume={27},
  number={5},
  pages={988--999},
  year={2024},
  publisher={Nature Publishing Group US New York}
}

@article{lynch2007frailty,
  title={The frailty of adaptive hypotheses for the origins of organismal complexity},
  author={Lynch, Michael},
  journal={Proceedings of the National Academy of Sciences},
  volume={104},
  number={suppl\_1},
  pages={8597--8604},
  year={2007},
  publisher={National Academy of Sciences}
}

@article{silver2021reward,
  title={Reward is enough},
  author={Silver, David and Singh, Satinder and Precup, Doina and Sutton, Richard S},
  journal={Artificial intelligence},
  volume={299},
  pages={103535},
  year={2021},
  publisher={Elsevier}
}

@book{marken2021study,
  title={The study of living control systems: A guide to doing research on purpose},
  author={Marken, Richard S},
  year={2021},
  publisher={Cambridge University Press}
}

@article{moreno2026intrinsic,
  title={How Intrinsic Motivation Underlies Embodied Open-Ended Behavior},
  author={Moreno-Bote, Rub{\'e}n and Haefner, Ralf and Galiano-Landeira, Jordi and Yang, Tianming and Maldonado, Pedro},
  journal={arXiv preprint arXiv:2601.10276},
  year={2026}
}

@article{henighan2020scaling,
  title={Scaling laws for autoregressive generative modeling},
  author={Henighan, Tom and Kaplan, Jared and Katz, Mor and Chen, Mark and Hesse, Christopher and Jackson, Jacob and Jun, Heewoo and Brown, Tom B and Dhariwal, Prafulla and Gray, Scott and others},
  journal={arXiv preprint arXiv:2010.14701},
  year={2020}
}

@article{Kaplan2020-scaling,
  author  = {Kaplan, Jared and McCandlish, Sam and Henighan, Tom
             and Brown, Tom B. and Chess, Benjamin and Child, Rewon
             and Gray, Scott and Radford, Alec and Wu, Jeffrey
             and Amodei, Dario},
  title   = {Scaling Laws for Neural Language Models},
  journal = {arXiv preprint arXiv:2001.08361},
  year    = {2020}
}

@inproceedings{Antonello2023-scaling,
  author    = {Antonello, Richard J. and Vaidya, Aditya R.
               and Huth, Alexander G.},
  title     = {Scaling Laws for Language Encoding Models in {fMRI}},
  booktitle = {Advances in Neural Information Processing Systems},
  volume    = {36},
  pages     = {21895--21907},
  year      = {2023}
}

@article{Grillner2018-limb,
  author  = {Grillner, Sten},
  title   = {Evolution: Vertebrate Limb Control over 420 Million Years},
  journal = {Current Biology},
  year    = {2018},
  volume  = {28},
  number  = {4},
  pages   = {R162--R164},
  doi     = {10.1016/j.cub.2017.12.040}
}

@article{Katz2013-circuits,
  author  = {Katz, Paul and Grillner, Sten and Wilson, Rachel and Borst, Alexander
             and Greenspan, Ralph and Buzs{\'a}ki, Gy{\"o}rgy and Martin, Kevan
             and Marder, Eve and Kristan, William and Friedrich, Rainer
             and Chklovskii, Dmitri Mitya},
  title   = {Vertebrate versus invertebrate neural circuits},
  journal = {Current Biology},
  year    = {2013},
  volume  = {23},
  number  = {12},
  pages   = {R504--R506},
  doi     = {10.1016/j.cub.2013.05.039}
}

@article{Zhao2024-baaiworm,
  author  = {Zhao, Mengdi and Wang, Ning and Jiang, Xinrui and Ma, Xiaoyang
             and Ma, Haixin and He, Gan and Du, Kai and Ma, Lei
             and Huang, Tiejun},
  title   = {An integrative data-driven model simulating {C. elegans} brain,
             body and environment interactions},
  journal = {Nature Computational Science},
  year    = {2024},
  volume  = {4},
  pages   = {978--990},
  doi     = {10.1038/s43588-024-00738-w}
}

@article{Schlegel2024-multiconnectome,
  author  = {Schlegel, Philipp and Yin, Yijie and Bates, Alexander S.
             and Dorkenwald, Sven and Eichler, Katharina and Brooks, Paul
             and Han, Daniel S. and Gkantia, Marina and dos Santos, Marcia
             and Munnelly, Eva J. and Badalamente, Griffin
             and Serratosa Capdevila, Laia and Sane, Varun A.
             and Fragniere, Alexandra M. C. and Kiassat, Ladann
             and Pleijzier, Markus W. and St{\"u}rner, Tomke
             and Tamimi, Imaan F. M. and Dunne, Christopher R.
             and Salgarella, Irene and Javier, Alexandre and Fang, Siqi
             and Perlman, Eric and Kazimiers, Tom and Jagannathan, Sridhar R.
             and Matsliah, Arie and Sterling, Amy R. and Yu, Szi-Chieh
             and McKellar, Claire E. and {FlyWire Consortium}
             and Costa, Marta and Seung, H. Sebastian and Murthy, Mala
             and Hartenstein, Volker and Bock, Davi D.
             and Jefferis, Gregory S. X. E.},
  title   = {Whole-brain annotation and multi-connectome cell typing of {Drosophila}},
  journal = {Nature},
  year    = {2024},
  volume  = {634},
  pages   = {139--152},
  doi     = {10.1038/s41586-024-07686-5}
}

@article{Brunton2026-sphinx,
  author  = {Brunton, Bingni W. and Abe, Elliott T. T.
             and Hu, Lawrence Jianqiao and Tuthill, John C.},
  title   = {The digital sphinx: Can a worm brain control a fly body?},
  journal = {eLife},
  year    = {2026},
  doi     = {10.7554/eLife.111516.1},
  note    = {Reviewed Preprint}
}

@INPROCEEDINGS{Geadah2024-hv,
  title     = "Inferring system and optimal control parameters of
               closed-loop systems from partial observations",
  author    = "Geadah, Victor and Arbelaiz, Juncal and Ritz, Harrison and
               Daw, Nathaniel D and Cohen, Jonathan D and Pillow, Jonathan W",
  booktitle = "2024 IEEE 63rd Conference on Decision and Control (CDC)",
  publisher = "IEEE",
  pages     = "8006--8013",
  month     =  dec,
  year      =  2024,
  doi       = "10.1109/CDC56724.2024.10886179",
  url       = "https://ieeexplore.ieee.org/document/10886179/",
  eprint    = "2502.15014"
}

@INCOLLECTION{Gander2008-uu,
  title     = "Nonlinear convergence analysis for the parareal algorithm",
  author    = "Gander, Martin J and Hairer, Ernst",
  booktitle = "Lecture Notes in Computational Science and Engineering",
  publisher = "Springer Berlin Heidelberg",
  address   = "Berlin, Heidelberg",
  pages     = "45--56",
  series    = "Lecture notes in computational science and engineering",
  year      =  2008,
  url       = "http://dx.doi.org/10.1007/978-3-540-75199-1_4",
  doi       = "10.1007/978-3-540-75199-1\_4",
  issn      = "1439-7358,2197-7100"
}

@article{driscoll2024flexible,
  title={Flexible multitask computation in recurrent networks utilizes shared dynamical motifs},
  author={Driscoll, Laura N and Shenoy, Krishna and Sussillo, David},
  journal={Nature Neuroscience},
  volume={27},
  number={7},
  pages={1349--1363},
  year={2024},
}

@ARTICLE{Sun2025-lz,
  title         = "Speed always wins: A survey on efficient architectures for
                   large Language Models",
  author        = "Sun, Weigao and Hu, Jiaxi and Zhou, Yucheng and Du, Jusen and
                   Lan, Disen and Wang, Kexin and Zhu, Tong and Qu, Xiaoye and
                   Zhang, Yu and Mo, Xiaoyu and Liu, Daizong and Liang, Yuxuan
                   and Chen, Wenliang and Li, Guoqi and Cheng, Yu",
  journal       = "arXiv preprint arXiv:2508.09834",
  month         =  aug,
  year          =  2025,
  url           = "http://arxiv.org/abs/2508.09834",
  archivePrefix = "arXiv",
  primaryClass  = "cs.CL",
  eprint        = "2508.09834"
}

@INPROCEEDINGS{Smith2022-dy,
  title     = "Simplified State Space Layers for Sequence Modeling",
  author    = "Smith, Jimmy T H and Warrington, Andrew and Linderman, Scott",
  booktitle = "The Eleventh International Conference on Learning Representations",
  month     =  sep,
  year      =  2022,
  url       = "https://openreview.net/pdf?id=Ai8Hw3AXqks"
}

@inproceedings{Yang2024-vi,
  title         = "Parallelizing linear transformers with the delta rule over
                   sequence length",
  author        = "Yang, Songlin and Wang, Bailin and Zhang, Yu and Shen, Yikang
                   and Kim, Yoon",
  booktitle     = "Advances in Neural Information Processing Systems",
  year          =  2024
}

@INPROCEEDINGS{Gu2024-jj,
  title         = "Mamba: Linear-time sequence modeling with selective state
                   spaces",
  author        = "Gu, Albert and Dao, Tri",
  booktitle     = "Conference on Language Modeling (COLM)",
  month         =  oct,
  year          =  2024,
  url           = "https://openreview.net/forum?id=tEYskw1VY2",
  archivePrefix = "arXiv",
  primaryClass  = "cs.LG",
  eprint        = "2312.00752"
}

@ARTICLE{Linderman2019-ng,
  title    = "Hierarchical recurrent state space models reveal discrete and
              continuous dynamics of neural activity in {C}. elegans",
  author   = "Linderman, Scott and Nichols, Annika and Blei, David and Zimmer,
              Manuel and Paninski, Liam",
  journal  = "bioRxiv",
  pages    =  621540,
  month    =  apr,
  year     =  2019,
  url      = "https://www.biorxiv.org/content/10.1101/621540v1?rss=1",
  doi      = "10.1101/621540",
}

@ARTICLE{Kaifosh2025-rn,
  title     = "A generic non-invasive neuromotor interface for human-computer
               interaction",
  author    = "Kaifosh, Patrick and Reardon, Thomas R and {CTRL-labs at Reality
               Labs} and Allen, Brian D and Anderson, Chris and Arnoud, Sacha
               and Arora, Rahul and Atray, Mridu and Awad, Lana and Ayerbe,
               Francisco and Baker, Christopher and Baker, Nicholas and
               Barachant, Alexandre and Bard, Philip and Beltran, Wilman
               Pimentel and Berenzweig, Adam and Bhasin, Rohin and Bienkowski,
               Joe and Bittner, Sean and Boegner, Luke and Bolarinwa, Anu and
               Bosley, Don and Bracaglia, Matthew and Bräcklein, Mario and
               Brandwein, Maclyn and Bravate, Joe and Butler, Matt and Calhoun,
               Adam J and Chang, Chia-Jung and Chenet, Daniel and Chester,
               Joshua and Chiarito, Rudi and Chitnis, Rohan and Choi, John and
               Chun, Won and Chung, Jeremiah and Connors, James and Costa, Jota
               and Cramer, Mark and Cunningham, Raven and Cusack, William F and
               Danielson, Nathan and Davidson, Thomas J and De Araujo, Bruno and
               DiMaiolo, Bob and Draves, Scott and Du, Alan and Edelson, Zaina
               and Enemuo, Phina and Fahmi, Mina and Farsad, Nariman and
               Farshchian, Ali and Feliz, Randy and Fine, Jake and Formento,
               Emanuele and Freeman, Dustin and Fu, Jianing and Gagnon-Audet,
               Jean-Christophe and Gajurel, Rupesh and Gamutan, Jonathan and
               Gao, Sida and Garcia, Jonateal and Gayraud, Nathalie Therese
               Helene and Ghani, Minha and Ghosh, Sayan and Gidwani, Vickram and
               Giebisch, Danny and Gimler, Greg and Gramfort, Alexandre and
               Grosberg, Lauren and Gunther, Bryn and Guo, Ning and Gupta,
               Chetan and Kaya, Sinem Guven and Ha, Austin and Hadjer, Katarina
               and Hernández, Carlos Xavier and Hertz, Stav and Hewitt, Carl and
               Hill, Daniel N and Hong, Kirak and Hong, Lillian and Hou, Helen
               and Hruda, Stepan and Hsieh, Alex and Hsiung, Vivian and Huang,
               Rongqing and Hui, Yue and Hulet, Hazel and Islam, Shaker and
               Jayaram, Vinay and Jiang, Connie and Jiang, Xiaodong and Juarez,
               Brooke and Jun, James Jaeyoon and Jun, Na Young and Kadakia,
               Nirag and Kakar, Nishant and Kamdar, Ajay and Kao, Ta-Chu and
               Kober, Steven and Koh, T W and Koshy, Christina Shabu and Lawn,
               Andrzej and Lee, Claire and Lee, Jennifer and Lee, Jinhyung and
               Lee, Juheui Amy and Li, Tiffanie and Liao, Jonathan and Liu,
               Yingru and Liu, Yuxuan and Lively, Saar and London, Kati and
               Louie, Roddy and Luongo, Francisco and Maczak, Attila and
               Maheswaranathan, Niru and Mandel, Michael and Marshall, Jesse and
               Marshall, Najja and Martincik, Mirek and Masse, Nicolas Yvan and
               McAnearney, Stephen and McHugh, Ashley and Menendez, Jorge
               Aurelio and Merel, Josh and Miller, David and Milyavskiy, Ilya
               and Monti, Ricardo Pio and Moore, Sean and Morin, Yonathan and
               Morrell, Brock and Morrison, Dano and Moschella, Anthony and
               Mulumudi, Suman and Muth, Conner and Naik, Krunal and Nakagaki,
               Norris and Nathan, Ajay and Nelson, Romario and Ngeo, Jimson and
               Nguyen, Keven and O'Connor, Luke and Ohayon, Shay and Orchard,
               Garrick and Osborn, Chris and Otchy, Timothy M and Owolabi,
               Emmanuella and Packer, Adam M and Pailla, Tejaswy and Paredes,
               Julia and Parker, Sean and Peixoto, Diogo and Perez, Matias and
               Perez, Zavion and Piérard, Adrien and Plaza, Stephen M and
               Plotkin, Natalie and Pnevmatikakis, Eftychios and Pool, Brandon
               and Puri, Shanil and Rajani, Sunaina and Fuentes, Jose Ramirez
               and Rojas, Julian Ramos and Ranjan, Tanvi and Reardon, Devin and
               Reid, Jonathan and Reisman, Jason and Rivera, Lain Warawao Nemo
               Mora y and Rolotti, Sebi and Rosenkranz, Andrew and Roth, Ian and
               Roy, Likhon and Rubin, Ran and Rudnicki, Alexander and Russell,
               Sam and Russo, Abby and Sacra, James and Sadoughi, Amir and
               Salim, Roxanna and Savane, Aichatou and Schlager, Collin and
               Schwab, David and Seely, Jeffrey and Seltzer, Mike and Sergin,
               Nurettin Dorukhan and Shah, Ami and Shah, Anish and Shamash,
               Philip and Sharma, Vandita and Shen, Stephie and Shi, Kevin and
               Shiah, Olivia and Siahpoosh, Yasmin and Siddiqi, Noor and
               Simpson, Jeremy and Singh, Gagandip and Sivakumar, Viswanath and
               Smith, Jeff and Sofuoglu, Seyyid Emre and Song, Ivy Jiyoung and
               Springer, Morgan and Spurr, Adrian and Stefanini, Fabio and
               Stout, Connor and Strauss, Emanuel and Suresh, Swetha and Suri,
               Ananya and Sussillo, David and Tang, Ziyi and Tank, Vikram and
               Tannady, Jesslyn and Tapia, Aliqyan and Tasci, Tugce and
               Tesileanu, Tiberiu and Tiwari, Aman and Tiwari, Anoushka and
               Tong, Calvin and Tormis, Blizelle and Trabulsi, Julia and
               Tsering, Migmar and Urquhart, Kyle and Walkington, Peter and
               Wang, Megan and Wang, Renxiong and Wang, Zhuo and Warden, Christy
               and Warren, Richard and Warriner, Claire L and Weiss, Ron J and
               Wetmore, Daniel Z and White, Ezri and Wiebe, Christopher and
               Williams, Steve and Xing, Yuguan and Ye, Chris and Yembarwar,
               Akshay and Yuan, Shuibenyang and Zawadzki, Michael and Zhang,
               Mingrui and Zhao, Jiesi and Zheng, Kevin and Zhong, Joseph and
               Zhou, Lei and Zlobinsky, Danny",
  journal   = "Nature",
  publisher = "Springer Science and Business Media LLC",
  pages     = "1--10",
  month     =  jul,
  year      =  2025,
  url       = "http://dx.doi.org/10.1038/s41586-025-09255-w",
  doi       = "10.1038/s41586-025-09255-w",
  issn      = "0028-0836,1476-4687",
}

@ARTICLE{Cotler2023-kr,
  title         = "Analyzing populations of neural networks via dynamical model
                   embedding",
  author        = "Cotler, Jordan and Tai, Kai Sheng and Hernández, Felipe and
                   Elias, Blake and Sussillo, David",
  journal       = "arXiv preprint arXiv:2302.14078",
  month         =  feb,
  year          =  2023,
  url           = "http://arxiv.org/abs/2302.14078",
  archivePrefix = "arXiv",
  primaryClass  = "cs.LG",
  eprint        = "2302.14078"
}

@ARTICLE{Safaie2023-pq,
  title     = "Preserved neural dynamics across animals performing similar
               behaviour",
  author    = "Safaie, Mostafa and Chang, Joanna C and Park, Junchol and Miller,
               Lee E and Dudman, Joshua T and Perich, Matthew G and Gallego,
               Juan A",
  journal   = "Nature",
  publisher = "Springer Science and Business Media LLC",
  volume    =  623,
  number    =  7988,
  pages     = "765--771",
  month     =  nov,
  year      =  2023,
  url       = "https://scholar.google.com/citations?view_op=view_citation&hl=en&user=8ss6egkAAAAJ&sortby=pubdate&citation_for_view=8ss6egkAAAAJ:_Re3VWB3Y0AC",
  doi       = "10.1038/s41586-023-06714-0",
  pmc       = "PMC10665198",
  pmid      =  37938772,
  issn      = "0028-0836,1476-4687",
}

@ARTICLE{Gallego2020-lj,
  title    = "Long-term stability of cortical population dynamics underlying
              consistent behavior",
  author   = "Gallego, Juan A and Perich, Matthew G and Chowdhury, Raeed H and
              Solla, Sara A and Miller, Lee E",
  journal  = "Nature neuroscience",
  month    =  jan,
  year     =  2020,
  url      = "https://doi.org/10.1038/s41593-019-0555-4",
  doi      = "10.1038/s41593-019-0555-4",
  issn     = "1097-6256,1546-1726"
}

@ARTICLE{IBL2025-il,
  title     = "Reproducibility of in vivo electrophysiological measurements in
               mice",
  author    = "{International Brain Laboratory} and Banga, Kush and Benson,
               Julius and Bhagat, Jai and Biderman, Dan and Birman, Daniel and
               Bonacchi, Niccolò and Bruijns, Sebastian A and Buchanan, Kelly
               and Campbell, Robert A A and Carandini, Matteo and Chapuis,
               Gaelle A and Churchland, Anne K and Davatolhagh, M Felicia and
               Lee, Hyun Dong and Faulkner, Mayo and Gerçek, Berk and Hu, Fei
               and Huntenburg, Julia and Hurwitz, Cole Lincoln and Khanal, Anup
               and Krasniak, Christopher and Lau, Petrina and Langfield,
               Christopher and Mackenzie, Nancy and Meijer, Guido T and Miska,
               Nathaniel J and Mohammadi, Zeinab and Noel, Jean-Paul and
               Paninski, Liam and Pan-Vazquez, Alejandro and Rossant, Cyrille
               and Roth, Noam and Schartner, Michael and Socha, Karolina Z and
               Steinmetz, Nicholas A and Svoboda, Karel and Taheri, Marsa and
               Urai, Anne E and Wang, Shuqi and Wells, Miles and West, Steven J
               and Whiteway, Matthew R and Winter, Olivier and Witten, Ilana B
               and Zhang, Yizi",
  journal   = "eLife",
  publisher = "eLife Sciences Publications Limited",
  volume    =  13,
  month     =  may,
  year      =  2025,
  url       = "http://dx.doi.org/10.7554/eLife.100840",
  doi       = "10.7554/eLife.100840",
  pmc       = "PMC12068871",
  pmid      =  40354112,
  issn      = "2050-084X",
}

@ARTICLE{IBL2021-ii,
  title     = "Standardized and reproducible measurement of decision-making in
               mice",
  author    = "{The International Brain Laboratory} and Aguillon-Rodriguez,
               Valeria and Angelaki, Dora and Bayer, Hannah and Bonacchi,
               Niccolo and Carandini, Matteo and Cazettes, Fanny and Chapuis,
               Gaelle and Churchland, Anne K and Dan, Yang and Dewitt, Eric and
               Faulkner, Mayo and Forrest, Hamish and Haetzel, Laura and
               Häusser, Michael and Hofer, Sonja B and Hu, Fei and Khanal, Anup
               and Krasniak, Christopher and Laranjeira, Ines and Mainen,
               Zachary F and Meijer, Guido and Miska, Nathaniel J and
               Mrsic-Flogel, Thomas D and Murakami, Masayoshi and Noel,
               Jean-Paul and Pan-Vazquez, Alejandro and Rossant, Cyrille and
               Sanders, Joshua and Socha, Karolina and Terry, Rebecca and Urai,
               Anne E and Vergara, Hernando and Wells, Miles and Wilson,
               Christian J and Witten, Ilana B and Wool, Lauren E and Zador,
               Anthony M",
  journal   = "eLife",
  publisher = "eLife Sciences Publications, Ltd",
  volume    =  10,
  number    = "biorxiv;2020.01.17.909838v4",
  pages     = "e63711",
  month     =  may,
  year      =  2021,
  url       = "https://doi.org/10.7554/eLife.63711",
  doi       = "10.7554/eLife.63711",
  issn      = "2050-084X",
}

@ARTICLE{Gutierrez2013-hr,
  title    = "Multiple mechanisms switch an electrically coupled, synaptically
              inhibited neuron between competing rhythmic oscillators",
  author   = "Gutierrez, Gabrielle J and O'Leary, Timothy and Marder, Eve",
  journal  = "Neuron",
  volume   =  77,
  number   =  5,
  pages    = "845--858",
  month    =  mar,
  year     =  2013,
  url      = "http://dx.doi.org/10.1016/j.neuron.2013.01.016",
  doi      = "10.1016/j.neuron.2013.01.016",
  pmc      = "PMC3664401",
  pmid     =  23473315,
  issn     = "0896-6273,1097-4199",
}

@ARTICLE{Kiebel2008-db,
  title     = "Dynamic causal modelling for {EEG} and {MEG}",
  author    = "Kiebel, Stefan J and Garrido, Marta I and Moran, Rosalyn J and
               Friston, Karl J",
  journal   = "Cognitive neurodynamics",
  publisher = "Springer",
  volume    =  2,
  number    =  2,
  pages     = "121--136",
  month     =  jun,
  year      =  2008,
  url       = "http://dx.doi.org/10.1007/s11571-008-9038-0",
  doi       = "10.1007/s11571-008-9038-0",
  pmc       = "PMC2427062",
  pmid      =  19003479,
  issn      = "1871-4080,1871-4099",
}

@INPROCEEDINGS{Ye2023-za,
  title     = "Neural Data Transformer 2: Multi-context Pretraining for Neural
               Spiking Activity",
  author    = "Ye, Joel and Collinger, Jennifer L and Wehbe, Leila and Gaunt,
               Robert",
  booktitle = "Thirty-seventh Conference on Neural Information Processing
               Systems",
  month     =  nov,
  year      =  2023,
  url       = "https://openreview.net/pdf?id=CBBtMnlTGq"
}

@INPROCEEDINGS{Azabou2024-bs,
  title     = "Multi-session, multi-task neural decoding from distinct
               cell-types and brain regions",
  author    = "Azabou, Mehdi and Pan, Krystal Xuejing and Arora, Vinam and
               Knight, Ian Jarratt and Dyer, Eva L and Richards, Blake Aaron",
  booktitle = "The Thirteenth International Conference on Learning
               Representations",
  year      =  2024,
  url       = "https://openreview.net/forum?id=IuU0wcO0mo"
}

@INPROCEEDINGS{Azabou2023-on,
  title     = "A unified, scalable framework for neural population decoding",
  author    = "Azabou, Mehdi and Arora, Vinam and Ganesh, Venkataramana and Mao,
               Ximeng and Nachimuthu, Santosh and Mendelson, Michael J and
               Richards, Blake A and Perich, M and Lajoie, Guillaume and Dyer,
               Eva L",
  booktitle = "Advances in Neural Information Processing Systems",
  month     =  oct,
  year      =  2023,
  url       = "http://dx.doi.org/10.48550/arXiv.2310.16046",
  doi       = "10.48550/arXiv.2310.16046",
  pmid      =  37961743
}

@INPROCEEDINGS{Ryoo2025-hl,
  title         = "Generalizable, real-time neural decoding with hybrid
                   state-space models",
  author        = "Ryoo, Avery Hee-Woon and Krishna, Nanda H and Mao, Ximeng and
                   Azabou, Mehdi and Dyer, Eva L and Perich, Matthew G and
                   Lajoie, Guillaume",
  booktitle     = "Advances in Neural Information Processing Systems",
  month         =  dec,
  year          =  2025,
  url           = "https://openreview.net/forum?id=1i4wNFgHDd",
  archivePrefix = "arXiv",
  primaryClass  = "q-bio.NC",
  eprint        = "2506.05320"
}

@ARTICLE{van-Gelder1998,
  title     = "The dynamical hypothesis in cognitive science",
  author    = "van Gelder, Tim",
  journal   = "The behavioral and brain sciences",
  publisher = "cambridge.org",
  volume    =  21,
  pages     = "615--628",
  month     =  oct,
  year      =  1998,
  url       = "https://www.cambridge.org/core/journals/behavioral-and-brain-sciences/article/dynamical-hypothesis-in-cognitive-science/C121F1B65A534F3E7A27075EE489AD1E",
  doi       = "10.1017/S0140525X98481731",
  issn      = "0140-525X,1469-1825"
}

@ARTICLE{Bommasani2021,
  title         = "On the Opportunities and Risks of Foundation Models",
  author        = "Bommasani, Rishi and Hudson, Drew A and Adeli, Ehsan and
                   Altman, Russ and Arora, Simran and von Arx, Sydney and
                   Bernstein, Michael S and Bohg, Jeannette and Bosselut,
                   Antoine and Brunskill, Emma and Brynjolfsson, Erik and Buch,
                   Shyamal and Card, Dallas and Castellon, Rodrigo and
                   Chatterji, Niladri and Chen, Annie and Creel, Kathleen and
                   Davis, Jared Quincy and Demszky, Dora and Donahue, Chris and
                   Doumbouya, Moussa and Durmus, Esin and Ermon, Stefano and
                   Etchemendy, John and Ethayarajh, Kawin and Fei-Fei, Li and
                   Finn, Chelsea and Gale, Trevor and Gillespie, Lauren and
                   Goel, Karan and Goodman, Noah and Grossman, Shelby and Guha,
                   Neel and Hashimoto, Tatsunori and Henderson, Peter and
                   Hewitt, John and Ho, Daniel E and Hong, Jenny and Hsu, Kyle
                   and Huang, Jing and Icard, Thomas and Jain, Saahil and
                   Jurafsky, Dan and Kalluri, Pratyusha and Karamcheti,
                   Siddharth and Keeling, Geoff and Khani, Fereshte and Khattab,
                   Omar and Koh, Pang Wei and Krass, Mark and Krishna, Ranjay
                   and Kuditipudi, Rohith and Kumar, Ananya and Ladhak, Faisal
                   and Lee, Mina and Lee, Tony and Leskovec, Jure and Levent,
                   Isabelle and Li, Xiang Lisa and Li, Xuechen and Ma, Tengyu
                   and Malik, Ali and Manning, Christopher D and Mirchandani,
                   Suvir and Mitchell, Eric and Munyikwa, Zanele and Nair, Suraj
                   and Narayan, Avanika and Narayanan, Deepak and Newman, Ben
                   and Nie, Allen and Niebles, Juan Carlos and Nilforoshan,
                   Hamed and Nyarko, Julian and Ogut, Giray and Orr, Laurel and
                   Papadimitriou, Isabel and Park, Joon Sung and Piech, Chris
                   and Portelance, Eva and Potts, Christopher and Raghunathan,
                   Aditi and Reich, Rob and Ren, Hongyu and Rong, Frieda and
                   Roohani, Yusuf and Ruiz, Camilo and Ryan, Jack and Ré,
                   Christopher and Sadigh, Dorsa and Sagawa, Shiori and
                   Santhanam, Keshav and Shih, Andy and Srinivasan, Krishnan and
                   Tamkin, Alex and Taori, Rohan and Thomas, Armin W and Tramèr,
                   Florian and Wang, Rose E and Wang, William and Wu, Bohan and
                   Wu, Jiajun and Wu, Yuhuai and Xie, Sang Michael and Yasunaga,
                   Michihiro and You, Jiaxuan and Zaharia, Matei and Zhang,
                   Michael and Zhang, Tianyi and Zhang, Xikun and Zhang, Yuhui
                   and Zheng, Lucia and Zhou, Kaitlyn and Liang, Percy",
  journal       = "arXiv preprint arXiv:2108.07258",
  month         =  aug,
  year          =  2021,
  url           = "https://arxiv.org/abs/2108.07258",
  archivePrefix = "arXiv",
  primaryClass  = "cs.LG",
  eprint        = "2108.07258"
}

@ARTICLE{Driscoll2018-lw,
  title    = "Computation through Cortical Dynamics",
  author   = "Driscoll, Laura N and Golub, Matthew D and Sussillo, David",
  journal  = "Neuron",
  volume   =  98,
  number   =  5,
  pages    = "873--875",
  month    =  jun,
  year     =  2018,
  url      = "http://dx.doi.org/10.1016/j.neuron.2018.05.029",
  doi      = "10.1016/j.neuron.2018.05.029",
  pmid     =  29879388,
  issn     = "0896-6273,1097-4199",
}

@BOOK{Rust2025-et,
  title     = "Elusive cures: Why neuroscience hasn’t solved brain disorders-and
               how we can change that",
  author    = "Rust, Nicole C",
  publisher = "Princeton University Press",
  month     =  dec,
  year      =  2025,
  url       = "https://www.degruyterbrill.com/document/doi/10.1515/9780691243078/html",
  doi       = "10.1515/9780691243078",
  isbn      =  9780691243078
}

@BOOK{Wiener1965-xv,
  title     = "Cybernetics Or Control and Communication in the Animal and the
               Machine",
  author    = "Wiener, Norbert",
  publisher = "MIT Press",
  year      =  1965,
  url       = "https://market.android.com/details?id=book-NnM-uISyywAC",
  isbn      =  9780262730099,
}

@Article{Little2012,
  author               = {Little, Simon and Brown, Peter},
  title                = {What brain signals are suitable for feedback control of deep brain stimulation in Parkinson's disease?},
  journal              = {Annals of the New York Academy of Sciences},
  year                 = {2012},
  volume               = {1265},
  number               = {1},
  pages                = {9--24},
  month                = aug,
  issn                 = {1749-6632},
  day                  = {1},
  doi                  = {10.1111/j.1749-6632.2012.06650.x},
  publisher            = {Blackwell Publishing Inc},
  url                  = {http://dx.doi.org/10.1111/j.1749-6632.2012.06650.x},
}

@article{Paninski2009,
    author = {Paninski, Liam and Ahmadian, Yashar and Ferreira, Daniel Gil G. and Koyama, Shinsuke and Rahnama Rad, Kamiar and Vidne, Michael and Vogelstein, Joshua and Wu, Wei},
    day = {1},
    doi = {10.1007/s10827-009-0179-x},
    issn = {1573-6873},
    journal = {Journal of computational neuroscience},
    month = aug,
    number = {1-2},
    pages = {107--126},
    pmcid = {PMC3712521},
    pmid = {19649698},
    publisher = {Springer Netherlands},
    title = {A new look at state-space models for neural data.},
    url = {http://dx.doi.org/10.1007/s10827-009-0179-x},
    volume = {29},
    year = {2010}
}

@article{ODoherty2011,
    author = {Hochberg, Leigh R. and Serruya, Mijail D. and Friehs, Gerhard M. and Mukand, Jon A. and Saleh, Maryam and Caplan, Abraham H. and Branner, Almut and Chen, David and Penn, Richard D. and Donoghue, John P.},
    day = {13},
    doi = {10.1038/nature04970},
    issn = {0028-0836},
    journal = {Nature},
    month = jul,
    number = {7099},
    pages = {164--171},
    pmid = {16838014},
    publisher = {Nature Publishing Group},
    title = {Neuronal ensemble control of prosthetic devices by a human with tetraplegia},
    url = {http://dx.doi.org/10.1038/nature04970},
    volume = {442},
    year = {2006}
}

@article{Kao2015b,
    author = {Kao, Jonathan C. and Nuyujukian, Paul and Ryu, Stephen I. and Churchland, Mark M. and Cunningham, John P. and Shenoy, Krishna V.},
    day = {29},
    doi = {10.1038/ncomms8759},
    issn = {2041-1723},
    journal = {Nature Communications},
    month = jul,
    pages = {7759+},
    title = {Single-trial dynamics of motor cortex and their applications to brain-machine interfaces},
    url = {http://dx.doi.org/10.1038/ncomms8759},
    volume = {6},
    year = {2015}
}

@article{Breakspear2017,
    author = {Breakspear, Michael},
    day = {23},
    doi = {10.1038/nn.4497},
    issn = {1097-6256},
    journal = {Nature Neuroscience},
    month = feb,
    number = {3},
    pages = {340--352},
    publisher = {Nature Research},
    title = {Dynamic models of large-scale brain activity},
    url = {http://dx.doi.org/10.1038/nn.4497},
    volume = {20},
    year = {2017}
}

@Article{sussillo2013opening,
	title={Opening the black box: low-dimensional dynamics in high-dimensional recurrent neural networks},
	author={Sussillo, David and Barak, Omri},
	journal={Neural computation},
	volume={25},
	number={3},
	pages={626--649},
	year={2013},
	publisher={MIT Press}
}

@ARTICLE{Willett2021-dp,
  title     = "High-performance brain-to-text communication via handwriting",
  author    = "Willett, Francis R and Avansino, Donald T and Hochberg, Leigh R
               and Henderson, Jaimie M and Shenoy, Krishna V",
  journal   = "Nature",
  publisher = "Nature Publishing Group",
  volume    =  593,
  number    =  7858,
  pages     = "249--254",
  month     =  may,
  year      =  2021,
  url       = "https://www.nature.com/articles/s41586-021-03506-2",
  issn      = "0028-0836",
  doi       = "10.1038/s41586-021-03506-2",
  pmc       = "pmc8163299"
}

@incollection{Roweis2001,
	Author = {Sam Roweis and Zoubin Ghahramani},
	Booktitle = {{K}alman filtering and neural networks},
	Editor = {Haykin, Simon},
	Isbn = {0-471-22154-6},
	Pages = {175--220},
	Publisher = {John Wiley \& Sons, Inc},
	Title = {Learning nonlinear dynamical systems using the expectation-maximization algorithm},
	Year = {2001}}

@ARTICLE{Pandarinath2018,
  title    = "Inferring single-trial neural population dynamics using
              sequential auto-encoders",
  author   = "Pandarinath, Chethan and O'Shea, Daniel J and Collins, Jasmine
              and Jozefowicz, Rafal and Stavisky, Sergey D and Kao, Jonathan C
              and Trautmann, Eric M and Kaufman, Matthew T and Ryu, Stephen I
              and Hochberg, Leigh R and Henderson, Jaimie M and Shenoy, Krishna
              V and Abbott, L F and Sussillo, David",
  journal  = "Nat. Methods",
  volume   =  15,
  number   =  10,
  pages    = "805--815",
  month    =  oct,
  year     =  2018,
}

@Article{Haykin1998,
  author               = {Haykin, S. and Principe, J.},
  title                = {Making sense of a complex world [chaotic events modeling]},
  journal              = {IEEE Signal Processing Magazine},
  year                 = {1998},
  volume               = {15},
  number               = {3},
  pages                = {66--81},
  month                = may,
  issn                 = {10535888},
  doi                  = {10.1109/79.671132},
  url                  = {http://dx.doi.org/10.1109/79.671132},
}

@article{Ecker2014b,
    author = {Ecker, Alexander S. and Berens, Philipp and Cotton, R. James and Subramaniyan, Manivannan and Denfield, George H. and Cadwell, Cathryn R. and Smirnakis, Stelios M. and Bethge, Matthias and Tolias, Andreas S.},
    doi = {10.1016/j.neuron.2014.02.006},
    issn = {08966273},
    journal = {Neuron},
    month = apr,
    number = {1},
    pages = {235--248},
    title = {State Dependence of Noise Correlations in Macaque Primary Visual Cortex},
    url = {http://dx.doi.org/10.1016/j.neuron.2014.02.006},
    volume = {82},
    year = {2014}
}

@book{Sarkka2013,
    author = {Simo S\"arkk\"a},
    isbn = {9781107619289},
    publisher = {Cambridge University Press},
    title = {Bayesian filtering and smoothing},
    url = {http://www.worldcat.org/isbn/9781107619289},
    year = {2013}
}

@article{Mante2013,
    author = {Mante, Valerio and Sussillo, David and Shenoy, Krishna V. and Newsome, William T.},
    day = {7},
    doi = {10.1038/nature12742},
    issn = {0028-0836},
    journal = {Nature},
    month = nov,
    number = {7474},
    pages = {78--84},
    publisher = {Nature Publishing Group, a division of Macmillan Publishers Limited. All Rights Reserved.},
    title = {Context-dependent computation by recurrent dynamics in prefrontal cortex},
    url = {http://dx.doi.org/10.1038/nature12742},
    volume = {503},
    year = {2013}
}

@article{Yu2009,
    author = {Yu, Byron M. and Cunningham, John P. and Santhanam, Gopal and Ryu, Stephen I. and Shenoy, Krishna V. and Sahani, Maneesh},
    day = {01},
    doi = {10.1152/jn.90941.2008},
    issn = {0022-3077},
    journal = {Journal of neurophysiology},
    month = jul,
    number = {1},
    pages = {614--635},
    pmid = {19357332},
    publisher = {American Physiological Society},
    title = {Gaussian-process factor analysis for low-dimensional single-trial analysis of neural population activity.},
    url = {http://dx.doi.org/10.1152/jn.90941.2008},
    volume = {102},
    year = {2009}
}

@ARTICLE{Smith2021,
  title     = "Reverse engineering recurrent neural networks with Jacobian
               switching linear dynamical systems",
  author    = "Smith, Jimmy and Linderman, Scott and Sussillo, David",
  journal   = "Advances in Neural Information Processing Systems",
  volume    =  34,
  month     =  dec,
  year      =  2021,
  url       = "https://proceedings.neurips.cc/paper/2021/hash/8b77b4b5156dc11dec152c6c71481565-Abstract.html",
  issn      = "1049--5258"
}

@INPROCEEDINGS{Karpowicz2024-qe,
  title    = "Few-shot Algorithms for {COnsistent} Neural decoding ({FALCON})
              benchmark",
  author   = "Karpowicz, Brianna M and Ye, Joel and Fan, Chaofei and
              Tostado-Marcos, Pablo and Rizzoglio, Fabio and Washington, Clay
              and Scodeler, Thiago and de Lucena, Diogo and Nason-Tomaszewski,
              Samuel R and Mender, Matthew J and Ma, Xuan and Arneodo, Ezequiel
              Matias and Hochberg, Leigh R and Chestek, Cynthia A and Henderson,
              Jaimie M and Gentner, Timothy Q and Gilja, Vikash and Miller, Lee
              E and Rouse, Adam G and Gaunt, Robert A and Collinger, Jennifer L
              and Pandarinath, Chethan",
  booktitle = "Advances in Neural Information Processing Systems, Datasets and
               Benchmarks Track",
  month    =  dec,
  year     =  2024,
  url      = "https://openreview.net/forum?id=FN02v4nD8y",
  doi      = "10.1101/2024.09.15.613126",
  pmc      = "PMC11429771",
  pmid     =  39345641,
  language = "en"
}

@INPROCEEDINGS{Sun2023-sa,
  title     = "{MABe22}: A Multi-Species Multi-Task Benchmark for Learned
               Representations of Behavior",
  author    = "Sun, Jennifer J and Marks, Markus and Ulmer, Andrew Wesley and
               Chakraborty, Dipam and Geuther, Brian and Hayes, Edward and Jia,
               Heng and Kumar, Vivek and Oleszko, Sebastian and Partridge,
               Zachary and Peelman, Milan and Robie, Alice and Schretter,
               Catherine E and Sheppard, Keith and Sun, Chao and Uttarwar, Param
               and Wagner, Julian Morgan and Werner, Erik and Parker, Joseph and
               Perona, Pietro and Yue, Yisong and Branson, Kristin and Kennedy,
               Ann",
  booktitle = "\textit{Proceedings of the 40th International Conference on
               Machine Learning}",
  publisher = "PRML",
  year      =  2023,
  url       = "https://proceedings.mlr.press/v202/sun23g.html"
}

@ARTICLE{Reed2022-tm,
  title    = "A Generalist Agent",
  author   = "Reed, Scott and Żołna, Konrad and Parisotto, Emilio and
              Colmenarejo, Sergio Gómez and Novikov, Alexander and Barth-Maron,
              Gabriel and Giménez, Mai and Sulsky, Yury and Kay, Jackie and
              Springenberg, Jost Tobias and Eccles, Tom and Bruce, Jake and
              Razavi, Ali and Edwards, Ashley and Heess, Nicolas and Chen,
              Yutian and Hadsell, Raia and Vinyals, Oriol and Bordbar, Mahyar
              and de Freitas, Nando",
  journal  = "Transactions on Machine Learning Research",
  month    =  aug,
  year     =  2022,
  url      = "https://openreview.net/pdf?id=1ikK0kHjvj"
}

@INPROCEEDINGS{Pei2021-kf,
  title     = "Neural Latents Benchmark '21: Evaluating latent variable models
               of neural population activity",
  author    = "Pei, Felix and Ye, Joel and Zoltowski, David and Wu, Anqi and
               Chowdhury, Raeed H and Sohn, Hansem and O'Doherty, Joseph E and
               Shenoy, Krishna V and Kaufman, Matthew T and Churchland, Mark and
               Jazayeri, Mehrdad and Miller, Lee E and Pillow, Jonathan and
               Park, Il Memming and Dyer, Eva L and Pandarinath, Chethan",
  booktitle = "Advances in Neural Information Processing Systems,
               Track on Datasets and Benchmarks",
  month     =  sep,
  year      =  2021,
  url       = "http://arxiv.org/abs/2109.04463"
}

@ARTICLE{Ba2014-fl,
  title   = "Do Deep Nets Really Need to be Deep?",
  author  = "Ba, Jimmy and Caruana, Rich",
  journal = "Advances in Neural Information Processing Systems",
  volume  =  27,
  year    =  2014,
  url     = "https://proceedings.neurips.cc/paper_files/paper/2014/file/ea8fcd92d59581717e06eb187f10666d-Paper.pdf"
}

@ARTICLE{Hinton2015-qe,
  title         = "Distilling the Knowledge in a Neural Network",
  author        = "Hinton, Geoffrey and Vinyals, Oriol and Dean, Jeff",
  journal       = "arXiv preprint arXiv:1503.02531",
  note          = "NIPS Deep Learning Workshop",
  month         =  mar,
  year          =  2015,
  url           = "http://arxiv.org/abs/1503.02531",
  archivePrefix = "arXiv",
  primaryClass  = "stat.ML",
  eprint        = "1503.02531"
}

@ARTICLE{Genkin2020-ps,
  title     = "Moving beyond generalization to accurate interpretation of
               flexible models",
  author    = "Genkin, Mikhail and Engel, Tatiana A",
  journal   = "Nature machine intelligence",
  publisher = "Springer Science and Business Media LLC",
  volume    =  2,
  number    =  11,
  pages     = "674--683",
  month     =  oct,
  year      =  2020,
  url       = "http://dx.doi.org/10.1038/s42256-020-00242-6",
  doi       = "10.1038/s42256-020-00242-6",
  issn      = "2522-5839,2522-5839",
  language  = "en"
}

@INPROCEEDINGS{Valente2022-hr,
  title     = "Extracting computational mechanisms from neural data using
               low-rank {RNNs}",
  author    = "Valente, Adrian and Pillow, Jonathan W and Ostojic, Srdjan",
  editor    = "Oh, Alice H and Agarwal, Alekh and Belgrave, Danielle and Cho,
               Kyunghyun",
  booktitle = "Advances in Neural Information Processing Systems",
  year      =  2022,
  url       = "https://openreview.net/forum?id=M12autRxeeS"
}

@ARTICLE{Fournier2023-ed,
  title     = "A practical survey on faster and lighter Transformers",
  author    = "Fournier, Quentin and Caron, Gaétan Marceau and Aloise, Daniel",
  journal   = "ACM computing surveys",
  publisher = "Association for Computing Machinery (ACM)",
  volume    =  55,
  number    = "14s",
  pages     = "1--40",
  month     =  dec,
  year      =  2023,
  url       = "http://dx.doi.org/10.1145/3586074",
  doi       = "10.1145/3586074",
  issn      = "0360-0300,1557-7341",
  language  = "en"
}

@INPROCEEDINGS{Bottou2007-ge,
  title     = "The tradeoffs of large scale learning",
  author    = "Bottou, Leon and Bousquet, Olivier",
  booktitle = "Advances in Neural Information Processing Systems",
  publisher = "Curran Associates, Inc.",
  year      =  2007,
  url       = "https://papers.nips.cc/paper/3323-the-tradeoffs-of-large-scale-learning"
}

@INPROCEEDINGS{Schaeffer2020-ib,
  title     = "Reverse-engineering recurrent neural network solutions to a
               hierarchical inference task for mice",
  author    = "Schaeffer, Rylan and Khona, Mikail and Meshulam, Leenoy and
               International, Brain Laboratory and Fiete, Ila",
  editor    = "Larochelle, H and Ranzato, M and Hadsell, R and Balcan, M F and
               Lin, H",
  booktitle = "Advances in Neural Information Processing Systems",
  publisher = "Curran Associates, Inc.",
  volume    =  33,
  pages     = "4584--4596",
  year      =  2020,
  url       = "https://proceedings.neurips.cc/paper_files/paper/2020/file/30f0641c041f03d94e95a76b9d8bd58f-Paper.pdf"
}

@ARTICLE{Haspel2023-if,
  title         = "The time is ripe to reverse engineer an entire nervous
                   system: simulating behavior from neural interactions",
  author        = "Haspel, Gal and Baker, Ben and Beets, Isabel and Boyden,
                   Edward S and Brown, Jeffrey and Church, George and Cohen,
                   Netta and Colon-Ramos, Daniel and Dyer, Eva and Fang-Yen,
                   Christopher and Flavell, Steven and Goodman, Miriam B and
                   Hart, Anne C and Izquierdo, Eduardo J and Kagias,
                   Konstantinos and Lockery, Shawn and Lu, Yangning and
                   Marblestone, Adam and Matelsky, Jordan and Mensh, Brett and
                   Pereira, Talmo D and Pfister, Hanspeter and Rajan, Kanaka and
                   Rotstein, Horacio G and Scholz, Monika and Shaevitz, Joshua W
                   and Shlizerman, Eli and Simeon, Quilee and Skuhersky, Michael
                   A and Tiruvadi, Vineet and Venkatachalam, Vivek and Wei,
                   Donglai and Wester, Brock and Yang, Guangyu Robert and
                   Yemini, Eviatar and Zimmer, Manuel and Kording, Konrad P",
  journal       = "arXiv preprint arXiv:2308.06578",
  month         =  aug,
  year          =  2023,
  url           = "http://arxiv.org/abs/2308.06578",
  archivePrefix = "arXiv",
  primaryClass  = "q-bio.NC",
  eprint        = "2308.06578"
}

@article{simeon2024homogenized,
  title={Homogenized $\textit{C. elegans}$ Neural Activity and Connectivity Data},
  author={Simeon, Quilee and Kashyap, Anshul and Kording, Konrad P and Boyden, Edward S},
  journal={arXiv preprint arXiv:2411.12091},
  year={2024}
}

@article{pagan2025individual,
  title={Individual variability of neural computations underlying flexible decisions},
  author={Pagan, Marino and Tang, Vincent D and Aoi, Mikio C and Pillow, Jonathan W and Mante, Valerio and Sussillo, David and Brody, Carlos D},
  journal={Nature},
  volume={639},
  number={8054},
  pages={421--429},
  year={2025},
  publisher={Nature Publishing Group UK London}
}

@BOOK{Ashby1952-lt,
  title  = "Design for a Brain",
  author = "Ross Ashby, W",
  publisher = "New York, Wiley",
  year   =  1952,
  url    = "https://www.ashby.info/Ashby%20-%20Design%20for%20a%20Brain%20-%20The%20Origin%20of%20Adaptive%20Behavior.pdf"
}

@INPROCEEDINGS{Xia2025-am,
  title     = "Inpainting the Neural Picture: Inferring Unrecorded Brain Area
               Dynamics from Multi-Animal Datasets",
  author    = "Xia, Ji and Zhang, Yizi and Wang, Shuqi and Allen, Genevera and
               Paninski, Liam and Hurwitz, Cole and Miller, Kenneth",
  booktitle = "Advances in Neural Information Processing Systems",
  year      =  2025,
  url       = "https://neurips.cc/virtual/2025/poster/117392",
}

@ARTICLE{Teeters2008-mv,
  title    = "Data sharing for computational neuroscience",
  author   = "Teeters, Jeffrey L and Harris, Kenneth D and Millman, K Jarrod and
              Olshausen, Bruno A and Sommer, Friedrich T",
  journal  = "Neuroinformatics",
  volume   =  6,
  number   =  1,
  pages    = "47--55",
  month    =  feb,
  year     =  2008,
  url      = "http://dx.doi.org/10.1007/s12021-008-9009-y",
  doi      = "10.1007/s12021-008-9009-y",
  pmid     =  18259695,
  issn     = "1539-2791,1559-0089",
  language = "en"
}

@ARTICLE{Yang2019-ja,
  title    = "Task representations in neural networks trained to perform many
              cognitive tasks",
  author   = "Yang, Guangyu Robert and Joglekar, Madhura R and Song, H Francis
              and Newsome, William T and Wang, Xiao-Jing",
  journal  = "Nature neuroscience",
  month    =  jan,
  year     =  2019,
  url      = "http://dx.doi.org/10.1038/s41593-018-0310-2",
  doi      = "10.1038/s41593-018-0310-2",
  pmid     =  30643294,
  issn     = "1097-6256,1546-1726",
  language = "en"
}

@ARTICLE{Michaels2020-xs,
  title    = "A goal-driven modular neural network predicts parietofrontal
              neural dynamics during grasping",
  author   = "Michaels, Jonathan A and Schaffelhofer, Stefan and Agudelo-Toro,
              Andres and Scherberger, Hansjörg",
  journal  = "Proceedings of the National Academy of Sciences of the United
              States of America",
  volume   =  117,
  number   =  50,
  pages    = "32124--32135",
  month    =  dec,
  year     =  2020,
  url      = "http://dx.doi.org/10.1073/pnas.2005087117",
  doi      = "10.1073/pnas.2005087117",
  pmc      = "PMC7749336",
  pmid     =  33257539,
  issn     = "0027-8424,1091-6490",
  language = "en"
}

@article{sarkka2020temporal,
  title={Temporal parallelization of Bayesian smoothers},
  author={S{\"a}rkk{\"a}, Simo and Garc{\'\i}a-Fern{\'a}ndez, {\'A}ngel F},
  journal={IEEE Transactions on Automatic Control},
  volume={66},
  number={1},
  pages={299--306},
  year={2020},
  publisher={IEEE}
}

@INPROCEEDINGS{Ross2011-aa,
  title     = "A Reduction of Imitation Learning and Structured Prediction to
               No-Regret Online Learning",
  author    = "Ross, Stephane and Gordon, Geoffrey and Bagnell, Drew",
  booktitle = "Proceedings of the Fourteenth International Conference on
               Artificial Intelligence and Statistics",
  publisher = "JMLR Workshop and Conference Proceedings",
  pages     = "627--635",
  month     =  jun,
  year      =  2011,
  url       = "https://proceedings.mlr.press/v15/ross11a.html",
  language  = "en"
}

@ARTICLE{Levine2020-aa,
  title         = "Offline reinforcement learning: Tutorial, review, and
                   perspectives on open problems",
  author        = "Levine, Sergey and Kumar, Aviral and Tucker, George and Fu,
                   Justin",
  journal       = "arXiv preprint arXiv:2005.01643",
  month         =  may,
  year          =  2020,
  url           = "http://arxiv.org/abs/2005.01643",
  archivePrefix = "arXiv",
  primaryClass  = "cs.LG",
  eprint        = "2005.01643"
}

@ARTICLE{Ijspeert2013-jq,
  title     = "Dynamical movement primitives: learning attractor models for
               motor behaviors",
  author    = "Ijspeert, Auke Jan and Nakanishi, Jun and Hoffmann, Heiko and
               Pastor, Peter and Schaal, Stefan",
  journal   = "Neural computation",
  publisher = "MIT Press",
  volume    =  25,
  number    =  2,
  pages     = "328--373",
  month     =  feb,
  year      =  2013,
  url       = "http://dx.doi.org/10.1162/NECO_a_00393",
  doi       = "10.1162/NECO\_a\_00393",
  pmid      =  23148415,
  issn      = "0899-7667,1530-888X",
  language  = "en"
}

@INPROCEEDINGS{Merel2020-ly,
  title         = "Deep neuroethology of a virtual rodent",
  author        = "Merel, Josh and Aldarondo, Diego and Marshall, Jesse and
                   Tassa, Yuval and Wayne, Greg and Ölveczky, Bence",
  booktitle     = "International Conference on Learning Representations (ICLR)",
  month         =  apr,
  year          =  2020,
  url           = "https://openreview.net/forum?id=SyxrxR4KPS",
  archivePrefix = "arXiv",
  primaryClass  = "q-bio.NC",
  eprint        = "1911.09451"
}

@ARTICLE{Vaxenburg2025-fr,
  title     = "Whole-body physics simulation of fruit fly locomotion",
  author    = "Vaxenburg, Roman and Siwanowicz, Igor and Merel, Josh and Robie,
               A and Morrow, Carmen M and Novati, Guido and Stefanidi, Zinovia
               and Both, Gert-Jan and Card, G and Reiser, Michael B and
               Botvinick, Matthew M and Branson, Kristin M and Tassa, Yuval and
               Turaga, Srinivas C",
  journal   = "Nature",
  publisher = "nature.com",
  volume    =  643,
  pages     = "1312--1320",
  month     =  apr,
  year      =  2025,
  url       = "http://dx.doi.org/10.1038/s41586-025-09029-4",
  doi       = "10.1038/s41586-025-09029-4",
  pmid      =  40267984,
  issn      = "0028-0836,1476-4687"
}

@misc{Sutton2019-bl,
  author       = {Sutton, Richard S.},
  title        = {The Bitter Lesson},
  howpublished = {\url{http://www.incompleteideas.net/IncIdeas/BitterLesson.html}},
  year         = {2019},
  month        = mar,
  url          = {http://www.incompleteideas.net/IncIdeas/BitterLesson.html}
}

\appendix
\setcounter{section}{0}
\setcounter{equation}{0}
\renewcommand{\thesection}{S\arabic{section}}
\renewcommand{\thefigure}{S\arabic{figure}}
\renewcommand{\thetable}{S\arabic{table}}
\renewcommand{\theequation}{S\arabic{equation}}

\end{document}